\documentclass[12pt]{iopart}

\usepackage{iopams}  

\usepackage{graphicx}
\usepackage{multirow} 
\usepackage{subfigure}
\usepackage{color}
\usepackage{booktabs}
\usepackage{array}
\newcommand{\PreserveBackslash}[1]{\let\temp=\\#1\let\\=\temp} %
\newcolumntype{C}[1]{>{\PreserveBackslash\centering}p{#1}}
\newcolumntype{R}[1]{>{\PreserveBackslash\raggedleft}p{#1}}
\newcolumntype{L}[1]{>{\PreserveBackslash\raggedright}p{#1}}
\newcolumntype{T}{>{\hfil$}p{\mylen}<{$\hfil}} 
\usepackage{makecell}

\def \Rey {\mbox{Re}}
\def \Ca  {\mbox{Ca}}
\def \We  {\mbox{We}}
\newcommand{\ie}{i.e. }

\newcommand{\bra}[1]{\langle #1\rangle} 

\begin{document}

\title[]{Suspensions of deformable particles in Poiseuille flows at finite inertia}

\author{Luigi Filippo Chiara$^1$, Marco Edoardo Rosti$^2$, Francesco Picano$^1$ and Luca Brandt$^3$}
\address{$^1$Department of Industrial Engineering, University of Padova, Via Venezia 1, 35131 Padova, Italy}
\address{$^2$Complex Fluids and Flows Unit, Okinawa Institute of Science and Technology Graduate University, 1919-1 Tancha, Onna-son, Okinawa 904-0495, Japan}
\address{$^3$Linn\'{e} Flow Centre and SeRC (Swedish e-Science Research Centre), KTH Mechanics, SE 100 44 Stockholm, Sweden}

\eads{\mailto{marco.rosti@oist.jp}}

\begin{abstract}
We analyze a suspension of deformable particles in a pressure-driven flow. The suspension is composed of neutrally buoyant initially spherical particles and a Newtonian carrier fluid, and the flow is solved by means of direct numerical simulations, using a fully Eulerian method based on a one-continuum formulation. The solid phase is modeled with an incompressible viscous hyperelastic constitutive relation, and the flow is characterized by three main dimensionless parameters, namely the solid volume fraction, the Reynolds and capillary numbers. The dependency of the effective viscosity on these three quantities is investigated to study the inertial effects on a suspension of deformable particles. It can be observed that the suspension has a shear-thinning behavior, and the reduction in effective viscosity for high shear rates is emphasized in denser configurations. The separate analysis of the Reynolds and capillary numbers reveal that the effective viscosity depends more on the capillary than on the Reynolds number. In addition, our simulations exhibit a consistent tendency for deformable particles to move towards the center of the channel, where the shear rate is low. This phenomenon is particularly marked for very dilute suspensions, where a whole region near the wall is empty of particles. Furthermore, when the volume fraction is increased this near-wall region is gradually occupied, because of higher mutual particle interactions. Deformability also plays an important role in the process. Indeed, at high capillary numbers, particles are more sensitive to shear rate variations and can modify their shape more easily to accommodate a greater number of particles in the central region of the channel. Finally, the total stress budgets show that the relative particle-induced stress contribution increases with the volume fraction and Reynolds number, and decreases with the particle deformability.
\end{abstract}

\vspace{2pc}

\maketitle

\section{Introduction}
\label{sec:Introduction}

Particles suspended in a carrier fluid can be encountered in several domains, such as biology and geophysics, but also biomechanical, pharmaceutical and cement industries. Pyroclastic flows from volcanoes, fluidized beds, the blood flow in veins and arteries, sedimentation in sea beds and slurry flows are other examples of flows where the knowledge of the particle dynamics is relevant. Currently, it is still hard to estimate the pressure drop needed to drive particle laden flows, while it can be precisely forecasted in single-phase flows as a function of the Reynolds number \cite{pope_2001a} and the properties of the wall surface (e.g.\  porosity \cite{breugem_boersma_uittenbogaard_2006a, rosti_cortelezzi_quadrio_2015a, rosti_brandt_pinelli_2018a} and elasticity \cite{rosti_brandt_2017a}). The additional complexity of multiphase flows is due to the presence of extra parameters such as the size, shape and elasticity of particles which become relevant. The density difference with respect to the carrier fluid and the solid volume fraction, denoted here with $\Phi$, are also crucial to describe their dynamics. In the present work, we analyze moderate dense suspensions where particles are deformable, fully resolving the fluid-structure interactions thanks to numerical simulations. In recent years, there has been a growth in the number of studies related to deformable objects, with the analysis of a single liquid droplet with constant surface tension, red blood cells enclosed by a biological membrane, and cells with stiff nuclei \cite{loewenberg_1998a, gao_hu_castaneda_2012a, freund_2014a, takeishi_rosti_imai_wada_brandt_2019a, alizad-banaei_loiseau_lashgari_brandt_2017a, rosti_brandt_2018a}. Here, we expand the literature of the field (in particular the work in Ref.~\cite{rosti_brandt_2018a}) by studying a full suspension of deformable particles and by considering the effect of finite inertia (although in the laminar regime), thus studying the interaction of elasticity and inertia for the first time. Furthermore, we study the problem in a Poiseuille flow, thus considering a non-uniform shear rate in the domain with the consequent particle migration and accumulation documented below.

Already in 1911, Einstein \cite{einstein_1911a} showed that the suspension viscosity increases linearly with the particle volume fraction $\Phi$, when focusing in the limit of vanishing inertia and for dilute suspensions. Batchelor \cite{batchelor_1977a} and Batchelor and Green \cite{batchelor_green_1972a} added a second-order correction in $\Phi$, but for higher volume fractions, the viscosity increases faster than a second-order polynomial \cite{stickel_powell_2005a}. In these cases, it is necessary to turn to empirical fits such as the Eilers fit \cite{ferrini_ercolani_de-cindio_nicodemo_nicolais_ranaudo_1979a, zarraga_hill_leighton-jr_2000a, singh_nott_2003a, kulkarni_morris_2008a}, which matches the rheology of rigid particle suspensions at small Reynolds numbers, for both low and high concentrations. When inertia becomes finite, deviations from the behavior predicted by the different empirical fits start to occur \cite{lin_peery_schowalter_1970a, kulkarni_morris_2008a}, an effect that can be related to an increase in the effective volume fraction at intermediate values of $\Phi$ \cite{picano_breugem_mitra_brandt_2013a}. Furthermore, at high volume fractions, once friction forces become relevant, other interesting phenomena can also be observed \cite{morris_2018a}.

The comprehension of the rheology of deformable objects has been a challenging task for many years \cite{roscoe_1967a}. In this case, the capillary number $\Ca$ (the ratio between viscous and elastic forces) is the non-dimensional parameter governing particle deformability. When $\Ca$ is low, particles easily recover the equilibrium shape, and deformations are small. These configurations are dominated by elastic forces. In 1932, Taylor first assumed small deformations and showed that, for small $\Phi$, the coefficient of the linear term in Einstein's relation is a function of the ratio between the particle and fluid viscosities. Using perturbative expansions, later analytical studies \cite{cox_1969a, frankel_acrivos_1970a, choi_schowalter_1975a, pal_2003a} tried to extend the result to higher orders in $\Phi$ and $\Ca$, similarly to what done by Batchelor for rigid particles. Only recently, different authors used high-fidelity numerical simulations to analyse such problem \cite{ii_gong_sugiyama_wu_huang_takagi_2012a, kruger_kaoui_harting_2014a, oliveira_cunha_2015a, srivastava_malipeddi_sarkar_2016a, matsunaga_imai_yamaguchi_ishikawa_2016a, rosti_brandt_2018a}.

The present work considers the inertial flow of viscous hyperelastic deformable particles, generally used to describe rubber-like substances. This class of constitutive relations can often show non-linear stress-strain curves and have a constitutive equation that is only a function of the current state of deformation. Moreover, if the work done by the stresses during a deformation process is dependent only on the initial and final configurations, the behavior of the material is path independent, and a stored strain energy function or elastic potential can be defined \cite{bonet_wood_1997a}. Many studies used this class of constitutive relations to describe particles, capsules, vesicles and even red blood cells \cite{sugiyama_ii_takeuchi_takagi_matsumoto_2011a, ii_sugiyama_takeuchi_takagi_matsumoto_2011a, ii_gong_sugiyama_wu_huang_takagi_2012a, villone_hulsen_anderson_maffettone_2014a, villone_greco_hulsen_maffettone_2016a}.

The objective of this paper is twofold: on one hand, we would like to have a better understanding of the physics involved in suspensions of deformable particles by considering the interactions of inertial and elastic effects, while on the other hand, we want to implement a robust numerical tool that can be applied in many applications involving elastic objects in inertial flows. Particle-resolved numerical simulations (full DNS) are becoming increasingly important for analyzing multiscale physical phenomena and providing detailed results, which can be exploited in modeling (i.e. turbulence or interaction modeling). In this context, we hope this work will open new pathways, so that more accurate and reliable closure equations will be derived in the near future. Another aspect to be highlighted is the intrinsic difficulty of the problem under investigation. First of all, the problem is a 4-way coupling interaction, meaning that every single object in the simulation can interact with its neighbors and with the surrounding fluid. Additionally, the resolution required to simulate accurately dense deformable particle suspensions is significantly higher than for rigid objects.

The obtained results are very promising. We show that suspensions of deformable particles exhibit a shear-thinning behavior that is much more sensitive to changes of $\Ca$ rather than variations in $\Rey$. In particular, deformable particles appear to move towards the center of the channel, and this phenomenon is particularly marked in dilute suspensions. Finally, our computations reveal that the more deformable particles tend to accumulate more in the central region of the channel, as they can modify their shape more easily.

\section{Formulation}
\label{sec:Formulation}

\subsection{Governing Equations}
\label{sec:Governing Equations}

We consider the flow of a suspension of deformable viscous hyperelastic particles in an incompressible Newtonian viscous fluid. The flow, as in the Poiseuille configuration, is driven by an imposed constant discharge, and the resulting pressure gradient in the stream-wise direction is thus computed accordingly. The number of particles in the domain is summarized by the total solid volume fraction $\Phi$, which is the ratio between the volume occupied by the solid phase and the total volume. The solid suspension is neutrally buoyant, i.e. it has the same density of the carrier fluid  $\rho^{s}=\rho^{f}=\rho $, as in many analyses for biological systems \cite{torii2008fluid, zhao2008fixed}. The unstressed reference shape of a particle is a sphere and no external body forces are applied on the continua.

The fluid and solid phase motion is governed by the conservation of momentum, and the incompressibility constraint is enforced in each phase
\begin{eqnarray}
\label{eq:NS_phases}
\frac{\partial u_i^f}{\partial t} + \frac{\partial u_i^f u_j^f}{\partial x_j} = \frac{1}{\rho} \frac{\partial \sigma_{ij}^f}{\partial x_j} \qquad &\textrm{if }& x \in \Omega^{f}, \\
\frac{\partial u_i^f}{\partial x_i} = 0 \qquad &\textrm{if }& x \in \Omega^{f}, \\
\frac{\partial u_i^s}{\partial t} + \frac{\partial u_i^s u_j^s}{\partial x_j} = \frac{1}{\rho} \frac{\partial \sigma_{ij}^s}{\partial x_j} \qquad &\textrm{if }& x \in \Omega^{s}, \\
\frac{\partial u_i^s}{\partial x_i} = 0 \qquad &\textrm{if }& x \in \Omega^{s},
\end{eqnarray}
where the superscripts $^{f}$ and $^{s}$ denote the fluid and solid phase fields. In the previous balance equations, $u_i$ indicates the \textit{i-th} component of the velocity vector, $\rho$ is the mass density and $\sigma_{ij}$ is the Cauchy stress tensor. The latter has a different form in the two phases and characterizes the solid and liquid behavior.

The kinematic and dynamic interactions between the fluid and solid phases are determined by enforcing the continuity of the velocity and traction force at the fluid-structure interface, namely
\begin{eqnarray}
\label{eq:BC_phases}
u_i^f = u_i^s \qquad &\textrm{on }& \partial \Omega^{f,s}, \label{eq:bc-v}\\
\sigma_{ij}^f n_j = \sigma_{ij}^s n_j \qquad &\textrm{on }& \partial \Omega^{f,s} \label{eq:bc-sigma},
\end{eqnarray}
where $n_j$ denotes the \textit{j-th} component of the normal vector at the interface.

To numerically solve the fluid-structure interaction, it is useful to introduce the so-called one-continuum formulation \cite{tryggvason_sussman_hussaini_2007a}, where only one set of equations is solved over the whole domain $ \Omega = \Omega^{f} \cup \Omega^{s} $. This is achieved by introducing a monolithic velocity vector field $u$ valid everywhere. We also introduce a local solid volume fraction function $\phi$, applying the volume averaging procedure \cite{takeuchi_yuki_ueyama_kajishima_2010a, quintard_whitaker_1994b}. The latter behaves as an indication function: it is zero in the fluid phase and equal to one in the solid phase. Due to the finite representation of our discrete grid, we have $0 < \phi < 1$ close to the fluid-solid interface. In this light, we could interpret $ \phi $ as a smoothed Heaviside function at the grid scale. In particular, the isoline $\phi = 0.5$ represents the interface. Moreover, if we integrate $\phi$ over the entire volume, we obtain the global solid volume fraction in the domain, $\Phi$.
\begin{figure}[ht]
	\centering
	{
	\setlength{\fboxsep}{0pt}%
	\fbox{\includegraphics[width=0.7\textwidth]{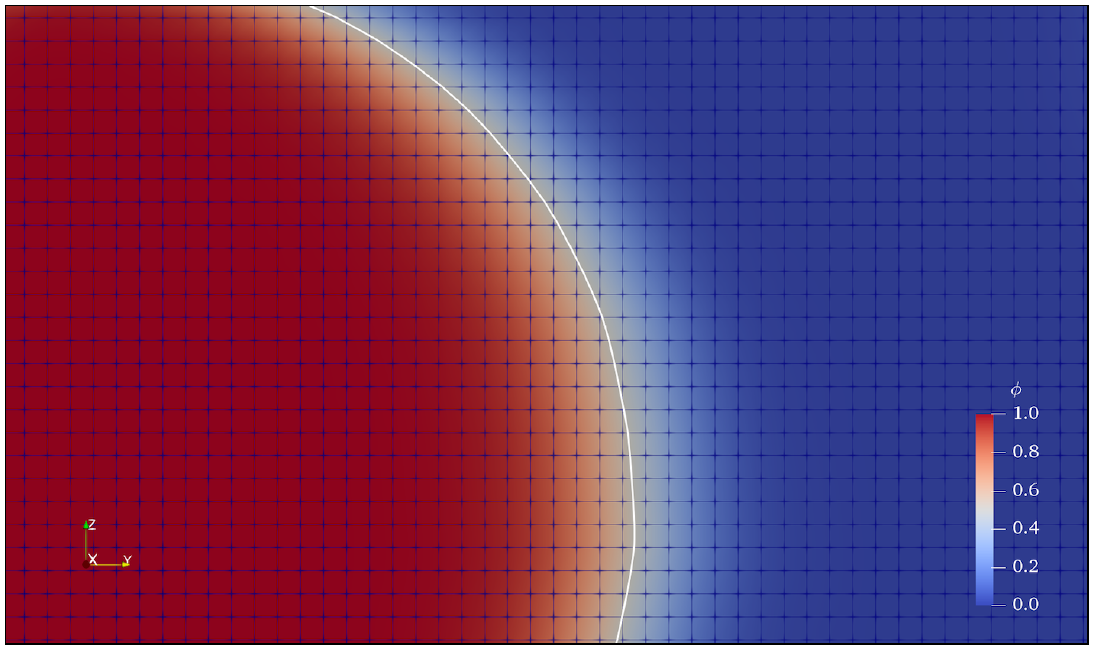}}
	}
	\caption{Zoomed view of the volume of fluid function $\phi$ at the grid level. The interface is located at the isoline $\phi = 0.5$, drawn in white in the picture. We can see that the transition between $\phi = 0$ (liquid phase, in blue) and $\phi = 1$ (solid phase, in red) is smooth. The resolution used, 48 grid points per particle diameter, is the same for every simulation.}
	\label{fig:vof_detail}
\end{figure}
The one-fluid formulation then reduces to
\begin{equation}
\label{eq:phi-stress}
\sigma_{ij} = \left( 1 - \phi \right) \sigma_{ij}^f + \phi \sigma_{ij}^s.
\end{equation}
The fluid is assumed to be Newtonian, so that the stress in the fluid phase can be written as
\begin{equation}
\label{eq:stress-f}
\sigma_{ij}^f = -p \delta_{ij} + 2 \mu^f D_{ij},
\end{equation}
where $p$ is the pressure, $\mu^f$ the fluid dynamic viscosity, $\delta_{ij}$ the Kronecker delta and $D_{ij}$ the strain rate tensor, defined as $D_{ij}= \left( \partial u_i/\partial x_j + \partial u_j/\partial x_i \right)/2$.
The solid is modeled as an incompressible viscous hyperelastic material, undergoing only the isochoric motion, with constitutive equation
\begin{equation}
\label{eq:stress-s-he}
\sigma_{ij}^s = -p \delta_{ij} + 2 \mu^s D_{ij} + G B_{ij},
\end{equation}
where the last term is the hyperelastic contribution modeled as a neo-Hookean material, thus satisfying the incompressible Mooney-Rivlin law. In equation~\ref{eq:stress-s-he}, $\mu^s$ is the solid dynamic viscosity, $B_{ij}$ the left Cauchy-Green deformation tensor, and $G$ is the modulus of transverse elasticity. In the following, we will assume that the solid and liquid phases have the same dynamic viscosity, so that $\mu^f = \mu^s$.

The set of equations for the solid material can be closed in a purely Eulerian manner by introducing a transport equation for the volume fraction $\phi$:
\begin{equation}
\label{eq:PHI-adv}
\frac{\partial \phi}{\partial t} + \frac{\partial u_k \phi}{\partial x_k} = 0,
\end{equation}
and updating the left Cauchy-Green deformation tensor with the following transport equation
\begin{equation}
\label{eq:B-adv}
\frac{\partial B_{ij}}{\partial t} + \frac{\partial u_k B_{ij}}{\partial x_k} = B_{kj}\frac{\partial u_i}{\partial x_k} + B_{ik}\frac{\partial u_j}{\partial x_k}.
\end{equation}

Combining all the previous equations we finally obtain
\begin{eqnarray}
\label{eq:Full_NS}
\rho \Big( \frac{\partial u_i}{\partial t} + \frac{\partial u_i u_j}{\partial x_j} \Big) = - \frac{\partial p}{\partial x_i} +  2 \mu^f \frac{\partial D_{ij}}{\partial x_j} + G \frac{\partial B_{ij}}{\partial x_j}, \\
\frac{\partial u_i}{\partial x_i} = 0,
\end{eqnarray}
where $B_{ij}$ is different from zero only in the presence of a solid particle. When $\Phi = 0$, the governing equations reduce to those for Newtonian fluids.



\begin{figure}
	\centering
	\subfigure[]{
	\label{fig:f_particles_a}
	\setlength{\fboxsep}{0pt}%
	\fbox{\includegraphics[width=0.45\textwidth]{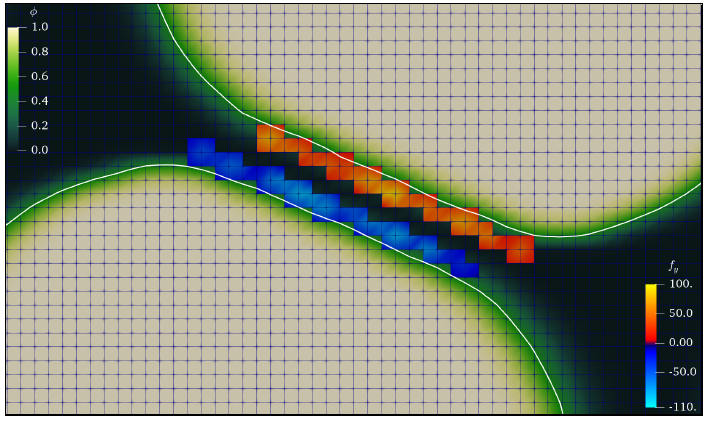}}
	}
	\subfigure[]{
	\label{fig:f_wall_b}
	\setlength{\fboxsep}{0pt}%
	\fbox{\includegraphics[width=0.45\textwidth]{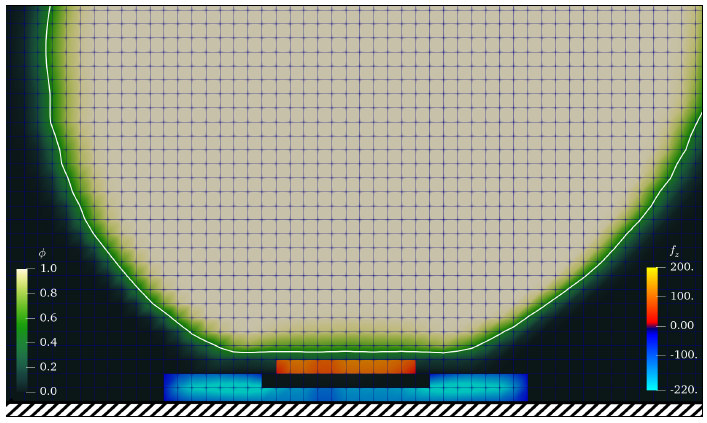}}
	}
	\caption{Intensity and stencil of the implemented interaction force. (a) $y$-component of the interaction force between two particles and (b) $z$-component of the interaction between a particle and the lower wall. In the figures, the positive components of the force are represented in orange, while the negative components are in blue. The forces are to be thought as applied at the center of each colored cell. Here, the volume fraction $\phi$ obeys a linear green color map, and the isoline $\phi=0.5$, in white, locates the interfaces.}
	\label{fig:interaction_force}
\end{figure}

The volume of fluid method in general does not include any repulsive force for particle-particle and particle-wall interactions. Without any kind of repulsive force, if two particles are close enough, they will eventually touch and merge. While this is possible for certain fluids in certain conditions (i.e. bubbles), we do not expect this behavior from solid particles where instead the solid stress will grow abrouptly. Thus, in order to avoid these anomalies, we introduce a subgrid repulsive force, as previously done in Ref.~\cite{bolotnov2011detached}. The following expression for the force is used \cite{clift1978bubbles}, both for particle-particle and particle-wall interactions:
\begin{equation}
\label{eq:interaction_force}
f_i = \mu^f V_b r \Big( \frac{a_1}{d} + \frac{a_2}{d^2} \Big) n_i,
\end{equation}
where $\mu^f$ is the fluid dynamic viscosity, $V_b$ the imposed bulk velocity, $r$ the particle radius, $d$ the distance between the two objects, $n_i$ the \textit{i-th} component of the local normal to the surface and $a_1 = 550.0$, $a_2 = 35.0$ are two model coefficients. This sub-grid force is only applied when two objects are close enough, i.e. $dx \leq d \leq 3dx$, where $dx$ is the uniform cell size used in the computations. The range and intensity of these interactions are displayed in figure~\ref{fig:interaction_force}. The force is always computed in the middle of the computational cells situated near the interface of two interacting particles, and directed towards the nearest surface. It is worth noting that these forces are applied only in the fluid phase and appear in the momentum balance as localized volume forces. They may also be interpreted as additional local pressure terms, which cannot be directly computed because of the grid resolution, which inevitably becomes inadequate when two particles are about to collide. The same approach has been used to prevent droplet coalescence in Refs.~\cite{fdv2019, de-vita_rosti_caserta_brandt_2020a}.

\subsection{Numerical implementation}
\label{sec:Numerical implementation}

In order to numerically simulate the flow, we follow the method for fluid-structure interaction problems first developed in Ref.~ \cite{sugiyama_ii_takeuchi_takagi_matsumoto_2011a}. 
Here, the time integration used to solve equations (\ref{eq:PHI-adv}), (\ref{eq:B-adv}), and (\ref{eq:Full_NS}) is based on an explicit fractional-step method \cite{kim_moin_1985a}. At each time step the volume fraction $\phi$ and left Cauchy-Green deformation tensor $B_{ij}$ are updated first, followed by the prediction step of the momentum conservation equations. All the terms are advanced in time with a low storage third-order Runge-Kutta scheme, except for the solid hyperelastic contribution, which is advanced with Crank-Nicolson, as previously done in \cite{min_yoo_choi_2001a}. The procedure is the same used in \cite{rosti_brandt_2018a, rosti_brandt_mitra_2018a}, where suspensions of deformable particle are analyzed in a Couette configuration. For further detail about the numerical implementation, the reader is referred to \cite{rosti_brandt_2017a}.


In order to implement the short-range interaction force, we need to be able to compute both local distances between particles and local normal directions. The continuous level set function is an answer to both problems. The level set function $\varphi$ is a scalar field defined in every point of the domain and 
its value is the (signed) distance from the nearest interface. According to this definition, the 0 contour surface of $\varphi$ represents exactly the solid-fluid interface.
%
%
%
As previously done in \cite{albadawi2013influence} among others, we reconstruct the level set function starting from the distribution of the volume of fluid function, at each time step. 
Once the level set function is computed, we are able to extract local distances and normal directions and numerically evaluate equation~\ref{eq:interaction_force}. More details on the reconstruction of the level set function and the implementation of the interaction force can be found in \cite{fdv2019}.


\subsection{Numerical Setup}
\label{sec:Numerical Setup}

We consider the flow of a Newtonian fluid laden with hyperelastic deformable spheres, and we investigate the final statistical flow configuration for different values of $\Phi$, $\Rey$ and $\Ca$.

The computational domain is a box of dimensions $3h \times 6h \times 2h$ in the span-wise, stream-wise and wall-normal directions, where $h$ is the half-channel height, our reference length. The initial shape of a particle is a sphere of radius $r = h/10$, as in \cite{picano_breugem_mitra_brandt_2013a}. The domain is subdivided into an uniform Cartesian mesh, with $48$ grid points per particle diameter, resulting in a computational box of $720 \times 1440 \times 480$ Eulerian grid points (for a total of almost 500 million points), in order to properly solve the interaction between the solid and fluid phases. No-slip boundary conditions are imposed on the solid upper and lower walls, while periodic boundary conditions are enforced in the homogeneous $x$ and $y$ directions.

In the following, three values of the channel bulk Reynolds number are considered, $\Rey_{b} = \rho h V_b/ \mu^f = $ 30, 250 and 500. For its definition we use the imposed bulk velocity in the channel $V_b$, while $\rho$ and $\mu^f$ are the density and the dynamic viscosity of the fluid phase. At each bulk Reynolds number corresponds a particle Reynolds number $\Rey = \rho \dot{\gamma} r^{2}/ \mu^f = $ 0.45, 3.75 and 7.5, where $\dot{\gamma}$ is the profile-averaged shear rate in the absence of particles, i.e. $\dot{\gamma} = 3 V_b/2h$. The range of variation of $\Rey$ was chosen comparable to what done for rigid particles in Ref.~\cite{picano_breugem_mitra_brandt_2013a}, where significant differences in the results had been observed.

The total solid volume fraction of the suspension $\Phi$ is defined as the volume average of the local volume fraction $\phi$, i.e. $\Phi=\bra{\phi}_{V}$. Hereafter, the double $\bra{\bra{\cdot}}$ indicates the time and volume average, $\bra{\cdot}_{V}$ the volume average (function of time) and the single $\bra{\cdot}$ the average in time, in the homogeneous $x$ and $y$ directions and in the lower and upper halves of the channel. Four values of total volume fraction $\Phi = 0.025$, $0.05$, $0.1$ and $0.15$ are considered, corresponding to a number of particles equal to $N =  215$, $430$, $859$ and $1289$.

As mentioned above, the capillary number $\Ca = \mu^f \dot{\gamma}/G$ is the main parameter governing the deformability of the suspension and represents the ratio between the viscous and elastic forces. Other possible deformability parameters are the modulus of transverse elasticity $G$ and the non-dimensional Weber number $\We =  \rho \dot{\gamma}^{2} r^2/G = \Rey \, \Ca $ (the ratio between the fluid inertia and the elastic forces). We study suspensions at four different capillary numbers, $\Ca =$ 0.009, 0.075, 0.15 and 0.3, in order to investigate the influence of the particle deformability on the suspension statistics. The smallest value corresponds to an almost rigid particle behavior, while the highest refers to considerably deformable particles.

\begin{table}
\centering
\setlength{\tabcolsep}{10pt}
\begin{tabular}{cc|c|c|c|c|}
& \multicolumn{1}{c}{} & \multicolumn{4}{ c }{$\Ca$} \\
\cline{3-6}
& & 0.009 & 0.075 & 0.15 & 0.3 \\
\cline{2-6}
\multicolumn{1}{ c|  }{} &
\multicolumn{1}{ c| }{7.5} & & & $\times$ & $\times$ \\
\cline{2-6}
\multicolumn{1}{ c|  }{$\Rey$} &
\multicolumn{1}{ c| }{3.75} & $\times$ & $\times$ & & $\times$ \\
\cline{2-6}
\multicolumn{1}{ c|  }{} &
\multicolumn{1}{ c| }{0.45} & $\times$ & & & $\times$ \\
\cline{2-6}
\end{tabular}
\caption{Different combinations of non-dimensional parameters under investigation. Three values of the Reynolds number are considered, together with four capillary numbers. For each of the 7 combinations of $\Rey$ (resp. $\Rey_b$) and $\Ca$ under examination (indicated with an $\times$ in the table), four volume fractions are considered, $\Phi = 0.025$, $0.05$, $0.1$ and $0.15$, resulting in a total of 28 numerical simulations. Thanks to this layout we have multiple simulations at constant $\Rey=3.75$, constant $\Ca=0.3$, and fixed $\Rey/\Ca$ ratio (diagonal).}
\label{tab:cases}
\end{table}

We run a total of 28 simulations, 7 for each volume fraction. The various combinations of parameters used are displayed in table~\ref{tab:cases}.
For all the cases, the solid viscosity is set equal to the fluid viscosity, $\mu^s = \mu^f$, and the same holds true for the mass densities $ \rho^{s} = \rho^{f} = \rho $.
All the simulations are started from the Poiseuille flow, a deterministic parabolic velocity profile along $z$ with a linear pressure distribution along the stream-wise direction $y$, together with a random distribution of the particles across the domain.
Each simulation reaches a statistical steady state after about 4 time units (\ie $t = 4 \ h/ V_b$). After a case reached its steady state, the calculations are continued for an interval of at least two time units, during which $30$ full flow fields are stored for further statistical analysis. To verify the convergence of the statistics, we computed them using different numbers of samples and verified that the differences are negligible.
Each case was run for 7 days using 512 cores, for a total of approximately $10^{5}$ CPU hours per case.

The present work is the natural continuation of \cite{rosti_brandt_2018a}, where the rheology of visco-elastic suspensions in a plane Couette flow is examined in the limit of vanishing inertia (at fixed $\Rey = 0.1$). Here, the authors studied different configurations letting $\Ca$ and $\Phi$ vary, with the particle material having the same constitutive equation introduced earlier in this paper. Also,
it is worth noting that the domain size is chosen to be the same as in a previous study \cite{lashgari_picano_breugem_brandt_2014a}, where inertial effects on rigid particle suspensions are studied.
Similarly to the box dimensions, also the general setup and most of the parameters used in this study are chosen as in \cite{lashgari_picano_breugem_brandt_2014a}
to ease comparisons.

\section{Results}
\label{sec:Results}

\begin{figure}[!h]
\centering
\includegraphics[width=0.9\textwidth]{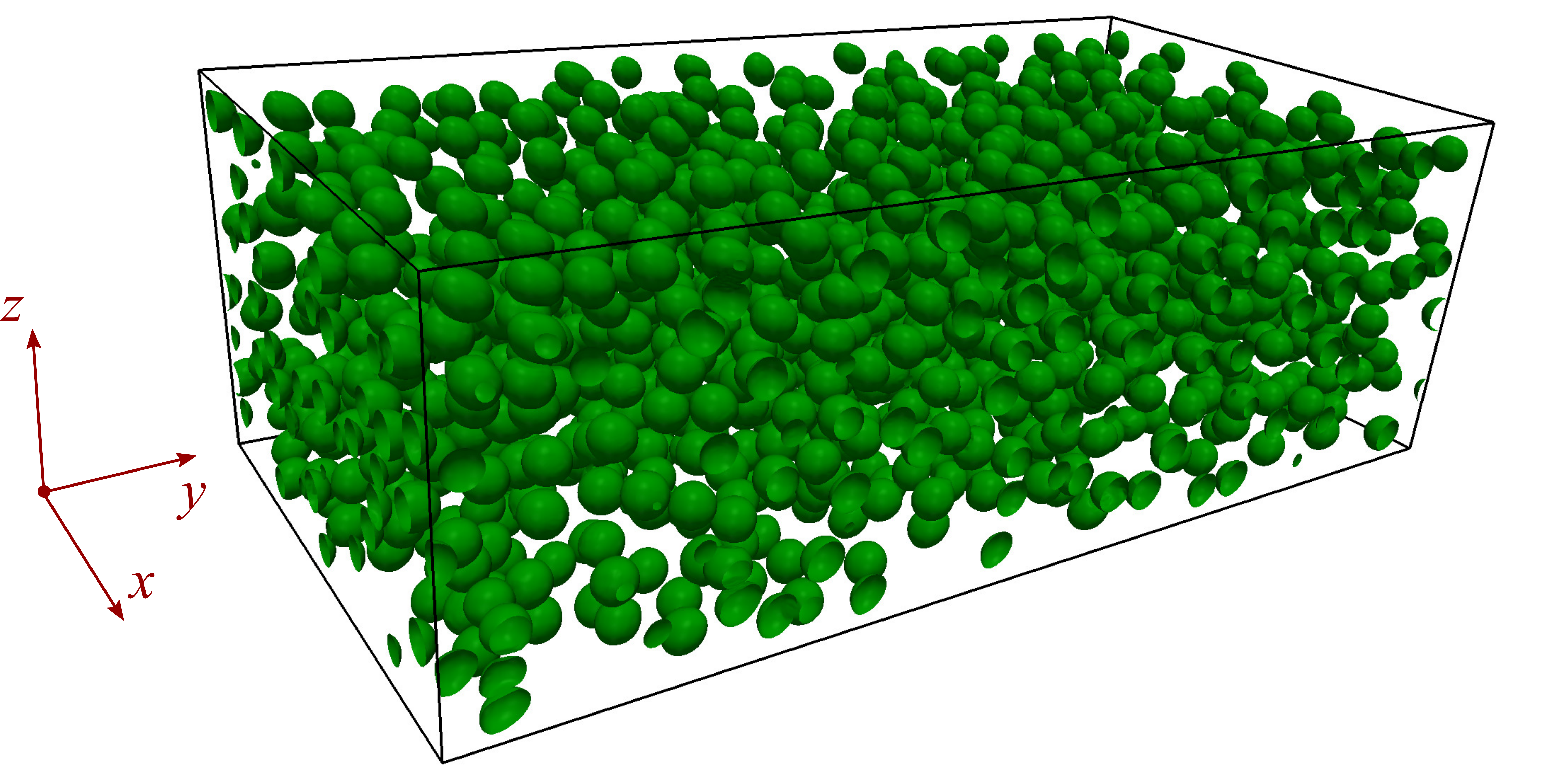}
\caption{Instantaneous particle arrangement in the $3h \times 6h \times 2h$ computational box, for a suspension at $\Phi = 0.15$, $\Rey = 3.75$ and $\Ca = 0.075$. The interfaces (in green) correspond to a $\phi = 0.5$ contour. The figure shows the coordinate system adopted in this manuscript, with $y$ indicating the stream-wise direction.}
\label{fig:palline}
\end{figure}

We first show in figure~\ref{fig:palline} a tridimensional visualization of the suspension at $\Phi = 0.15$, $\Rey = 3.75$ and $\Ca = 0.075$. It can be noticed how dense the suspension appears already at a volume fraction of $15\%$. As expected, the particle deformation is a function of the vertical coordinate and it is more evident near the walls where the local shear is higher.

\begin{figure}[!ht]
\centering
\includegraphics[width=0.7\textwidth]{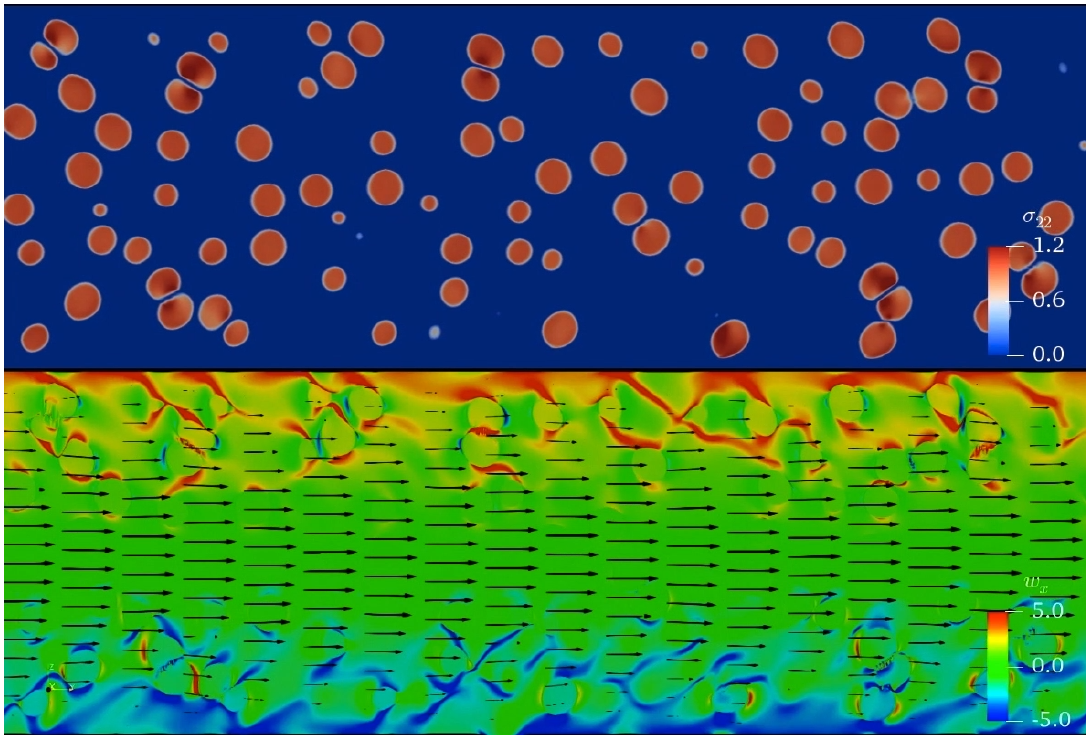}
\caption{Middle-plane stream-wise-wall-normal section of the suspension flow at $\Phi = 0.15$, $\Rey = 3.75$ and $\Ca = 0.075$. The panels illustrate (upper) the instantaneous stream-wise component of the hyperelastic shear stress and (lower) the span-wise component of the vorticity vector. The black arrows represent the projection of the velocity vector on the $y-z$ plane.}
\label{fig:visual}
\end{figure}

The typical behavior of the particle distribution and deformation for the same case can be observed looking at the upper panel of figure~\ref{fig:visual}, where the solid hyperelastic shear stress is shown in a stream-wise-wall-normal cut plane. The hyperelastic shear stress, always null (blue) outside the solid phase (red), appears high when two or more particles interact and are deformed from the unperturbed reference spherical shape. At this capillary number, the shear rate accounts for a modest deformation, which however becomes considerable near the walls where the velocity gradient is maximum. In the bottom panel of figure~\ref{fig:visual} the span-wise vorticity component is illustrated by contours together with the velocity field (black arrows) in the same plane as in the top panel. The velocity field appears close to the parabolic Poiseuille solution, with the addition of the perturbation induced locally by the particles. The vorticity appears intense near the wall and in the gap regions between interacting particles.


\begin{figure}[h!]
\subfigure[]{\includegraphics[width=0.5\textwidth]{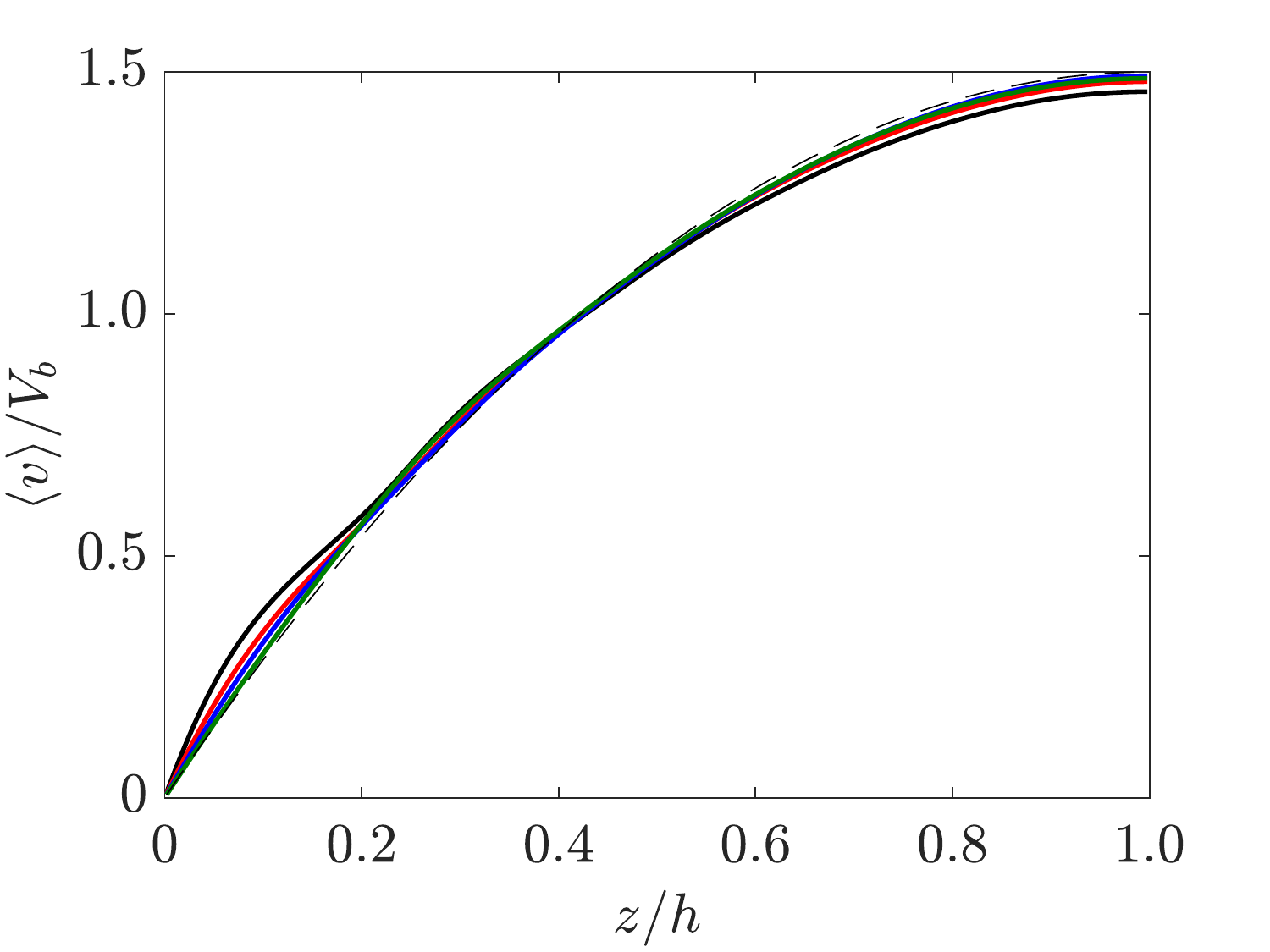} \label{fig:vel_mean_phi:a}}
\subfigure[]{\includegraphics[width=0.5\textwidth]{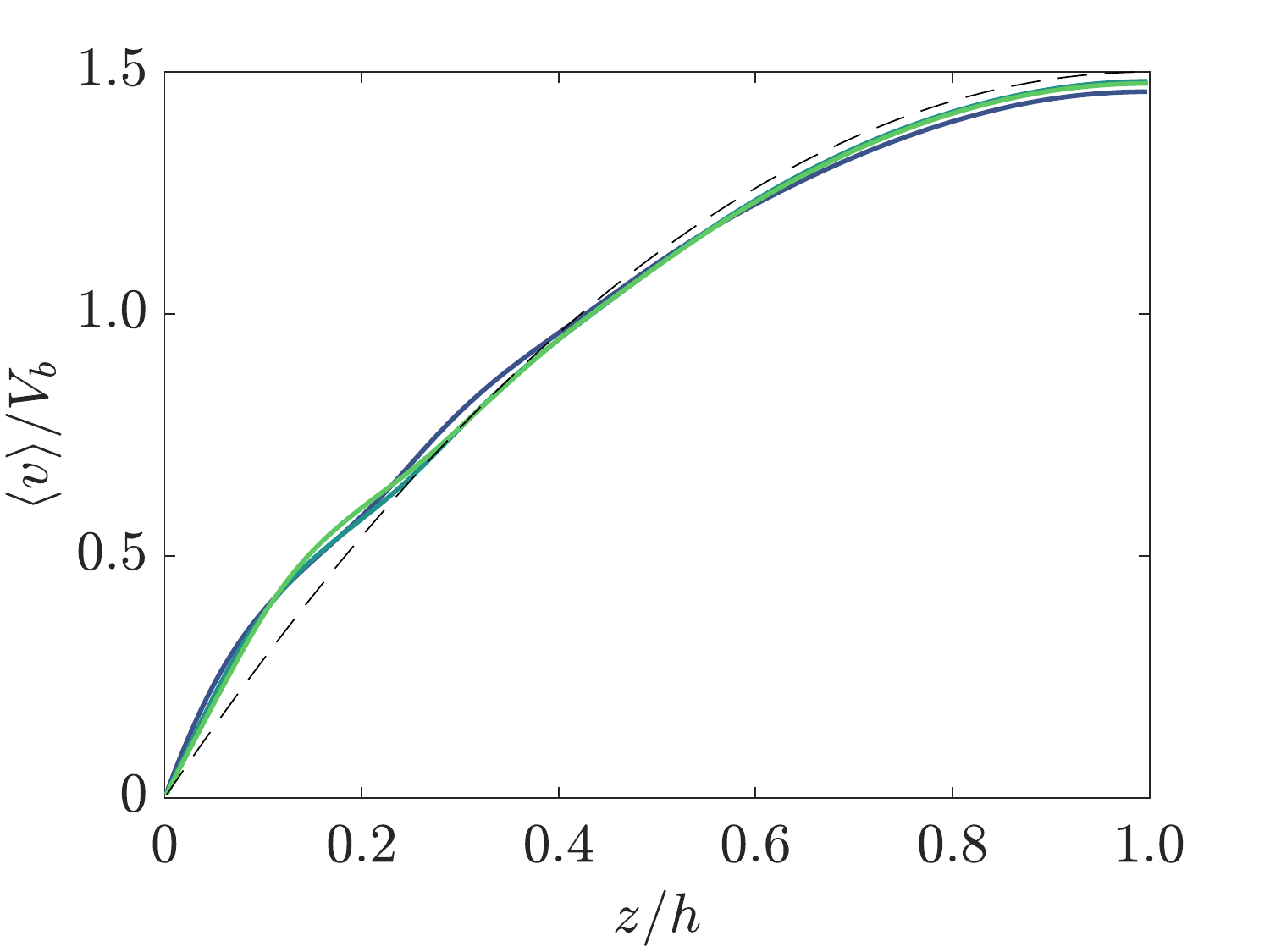} \label{fig:vel_mean_Ca:b}}
\caption{Stream-wise time-averaged velocity profiles for (a) different $\Phi$ at $\Rey = 3.75$ and $\Ca= 0.009$ and (b) different $\Ca$ at $\Rey = 3.75$ and $\Phi= 0.15$. On the left panel different colors represent different volume fractions: black, red, blue and green lines correspond to $\Phi = 0.15$, $0.1$, $0.05$ and $0.025$, respectively. On the right panel, data at $\Ca = 0.009$, $0.075$ and $0.3$ is plotted in blue, dark green and light green. As it can be seen, the velocity profiles are almost indistinguishable from each other and present a blunt shape typical of shear-thinning fluids.}
\label{fig:velocity_profiles}
\end{figure}

The mean stream-wise velocity profiles are plotted in figure~\ref{fig:velocity_profiles}. They do not differ significantly from the quadratic solution of the Navier-Stokes equation in the absence of particles, and show a slight shear-thinning behavior with a reduction of the centerline velocity and a slight increase near the walls.


\begin{figure}[h!]
\subfigure[]{\includegraphics[width=0.5\textwidth]{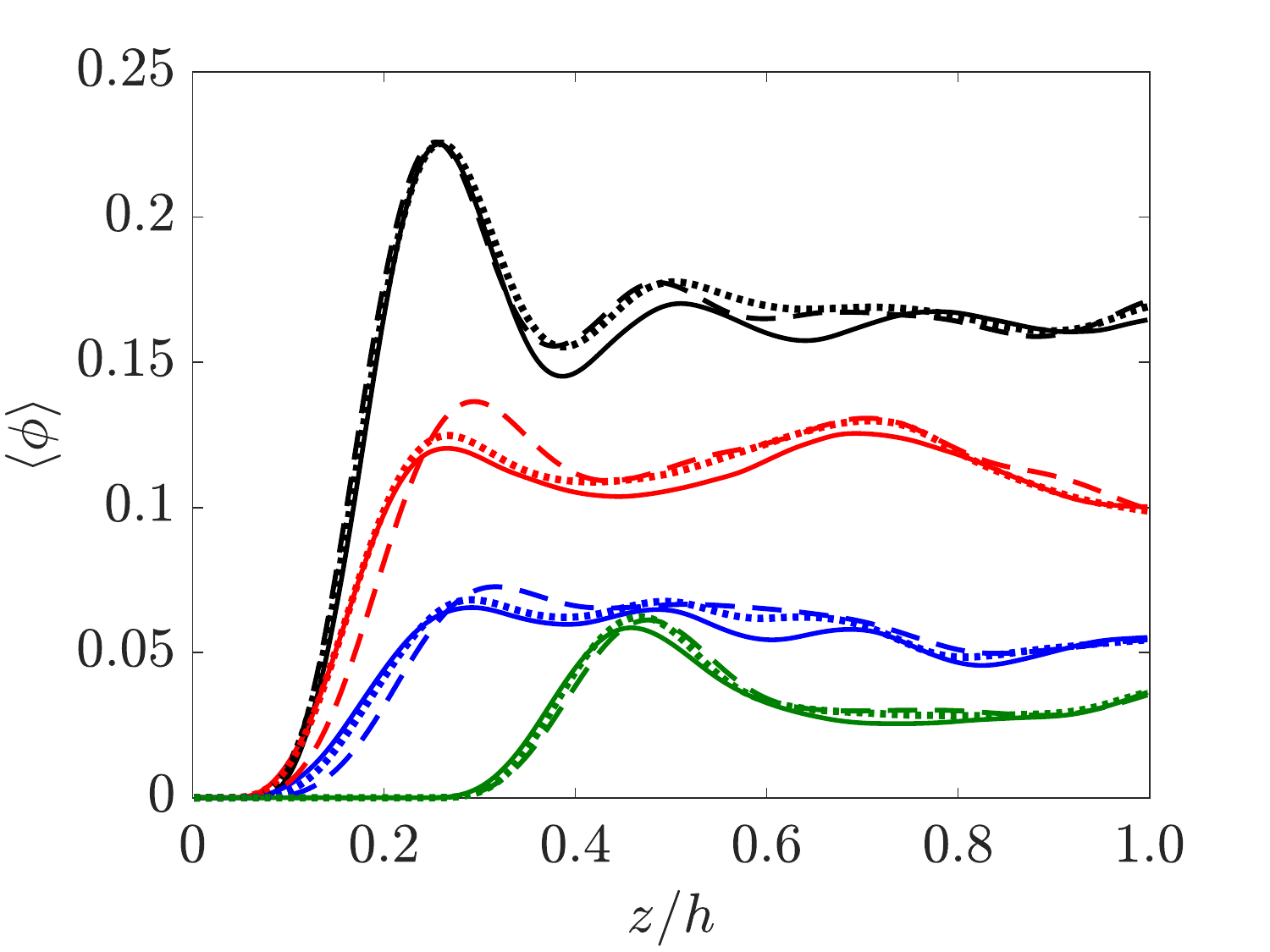}\label{fig:Phi_and_Re}}
\subfigure[]{\includegraphics[width=0.5\textwidth]{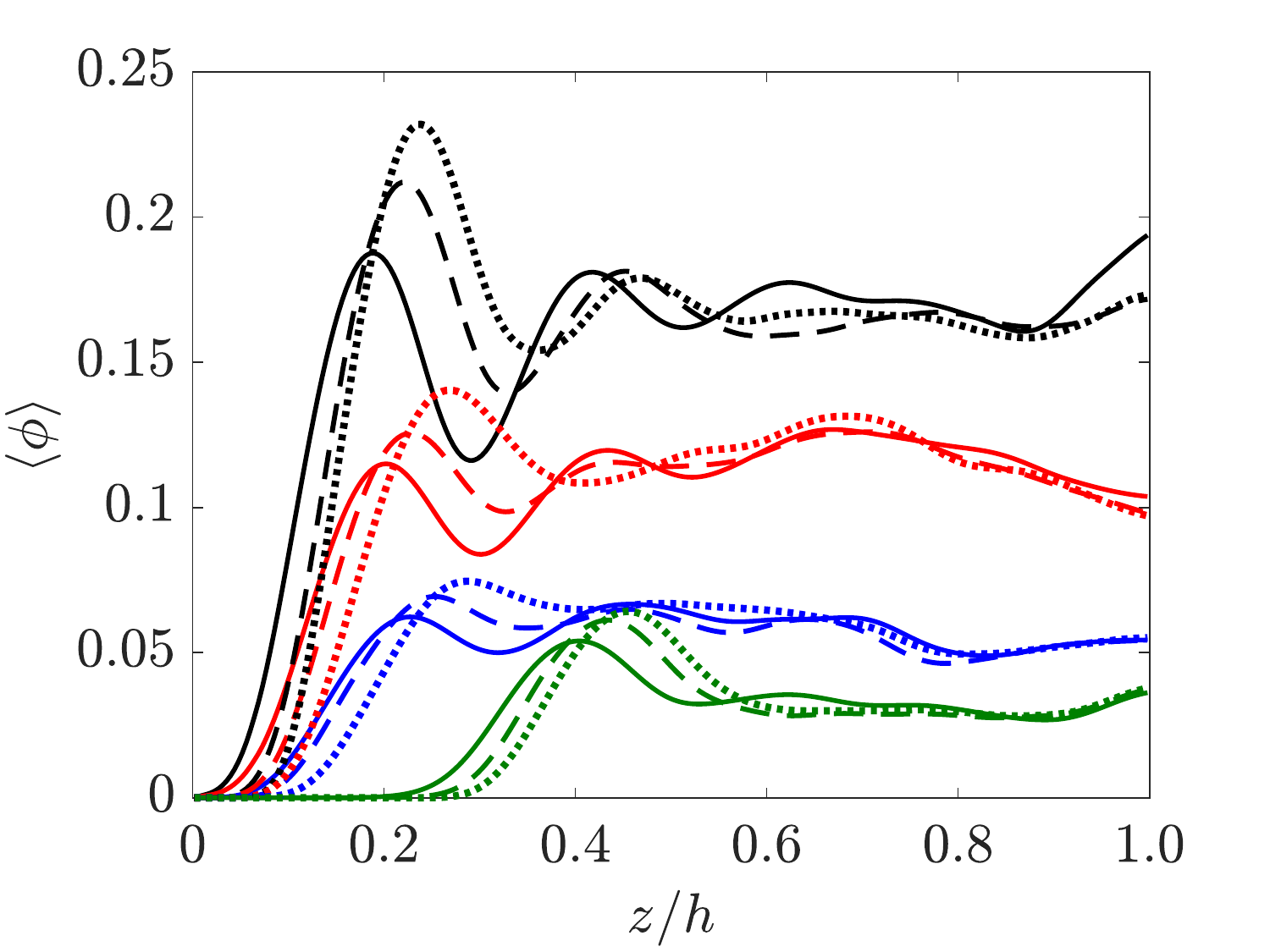}\label{fig:Phi_and_Ca}}
\caption{Averaged particle distributions $\langle \phi \rangle$ along the normalized wall-normal direction, for different total volume fractions, Reynolds and capillary numbers. The color scheme of both panels is the same of figure~\ref{fig:vel_mean_phi:a}, with black, red, blue and green lines indicating different volume fractions, i.e. $\Phi = 0.15$, $0.1$, $0.05$ and $0.025$. (a) Solid ---, dashed $--$, and dotted $\cdots$ lines represent data at $\Rey = 0.45$, $3.75$ and $7.5$, at a fixed $\Ca=0.3$. (b) Data at different capillary numbers, $\Ca = 0.009$ (---), $0.075$ ($--$) and $0.3$ ($\cdots$), at a fixed $\Rey=3.75$.}
\label{fig:Phi_over_z}
\end{figure}

We display the mean particle volume fraction $\langle \phi \rangle$ (averaged over $x$, $y$ and time) versus the wall-normal direction $z$ for $\Ca=0.3$ and different Reynolds numbers in figure~\ref{fig:Phi_and_Re}. It can be noted that for the smallest volume fraction considered here, e.g. $\Phi=0.025$, particles avoid the near-wall region, which appears depleted. Indeed, single deformable particles are known to migrate towards the pipe/channel center in Poiseuille flows, away from high shear regions \cite{dhiya2019}. In dilute cases, particle mutual interactions are rare and particles can migrate towards the channel center more freely, resulting in an empty portion of the channel near the wall.
On the contrary, the volume fraction profiles of denser configurations, i.e. $\Phi \ge 0.05$, show that the particles are distributed through the whole channel, with a clear layer near the wall. This is due to the mutual interactions between particles, increasing in number as $\Phi$ increases, forcing them to occupy the whole channel.
In the figure, it is also evident that the maximum of the volume fraction profiles, denoting wall layering, moves closer to the wall as $\Phi$ increases because of the augmented mutual interaction between the particles. The particle distribution appears weakly affected by changes in the Reynolds number in the range considered here. A near-wall region depleted of particles is noteworthy. Indeed, this result implies that, not only the wall-region has to be properly modeled to account for particle layering, but also the bulk of the channel can have a non-uniform distribution of particles, thus making the problem fully dependent not only on the local shear-rate as for rigid particles, but also on the local volume fraction.

Figure~\ref{fig:Phi_and_Ca} illustrates the local mean volume fraction at a fixed $\Rey=3.75$ and different $\Ca$. As $\Ca$ increases, it can be seen that the particles tend to show a peak slightly further away from the wall. This is because the more deformable particles are more sensitive to the shear rate variations (shear-induced migration), and they can modify their shape more easily to accommodate more particles in the channel center, thus leaving empty a bigger portion of the channel near the wall.


\begin{figure}[h!]
\subfigure[]{\includegraphics[width=0.5\textwidth]{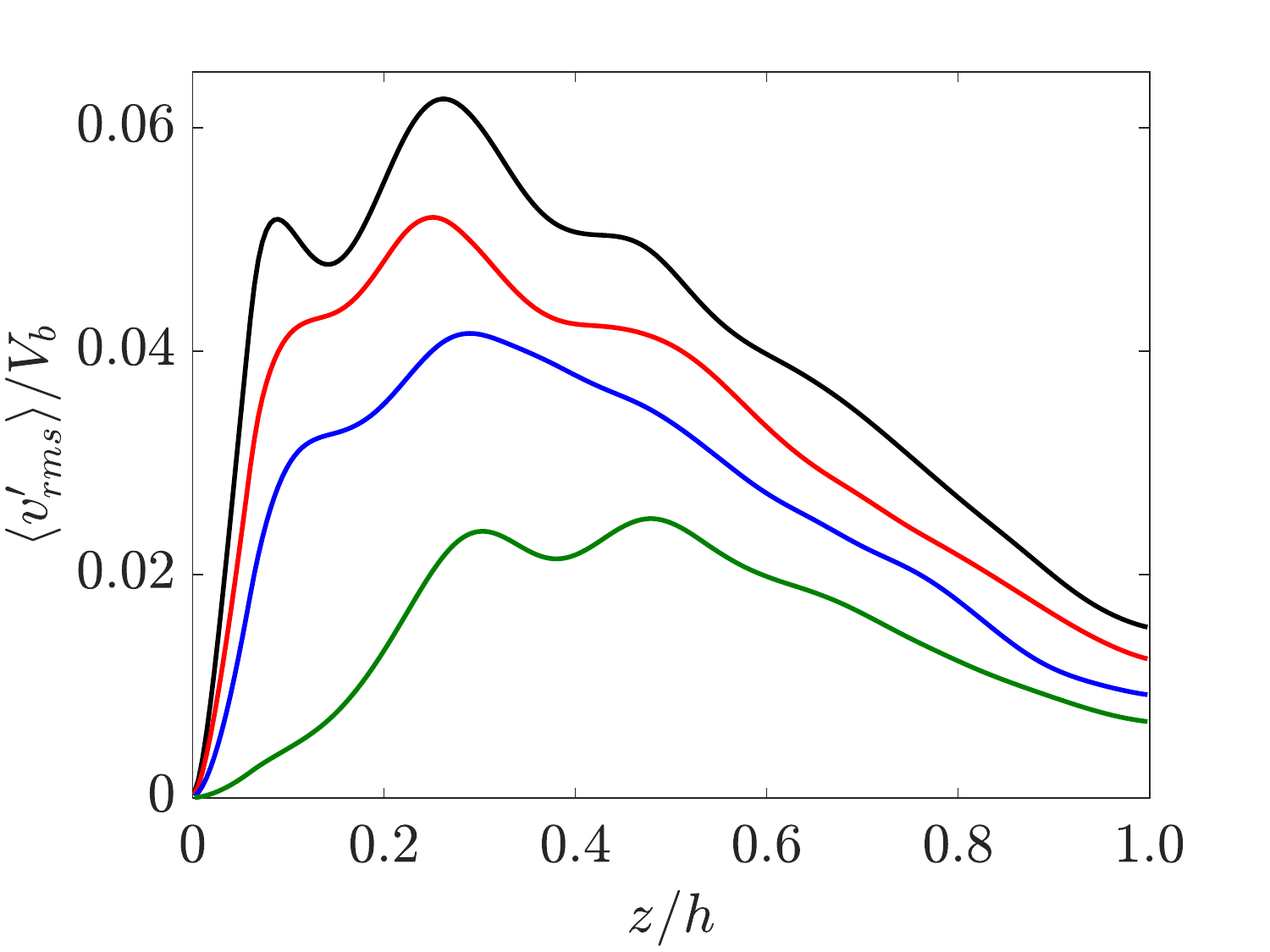} \label{fig:v_rms:a}}
\subfigure[]{\includegraphics[width=0.5\textwidth]{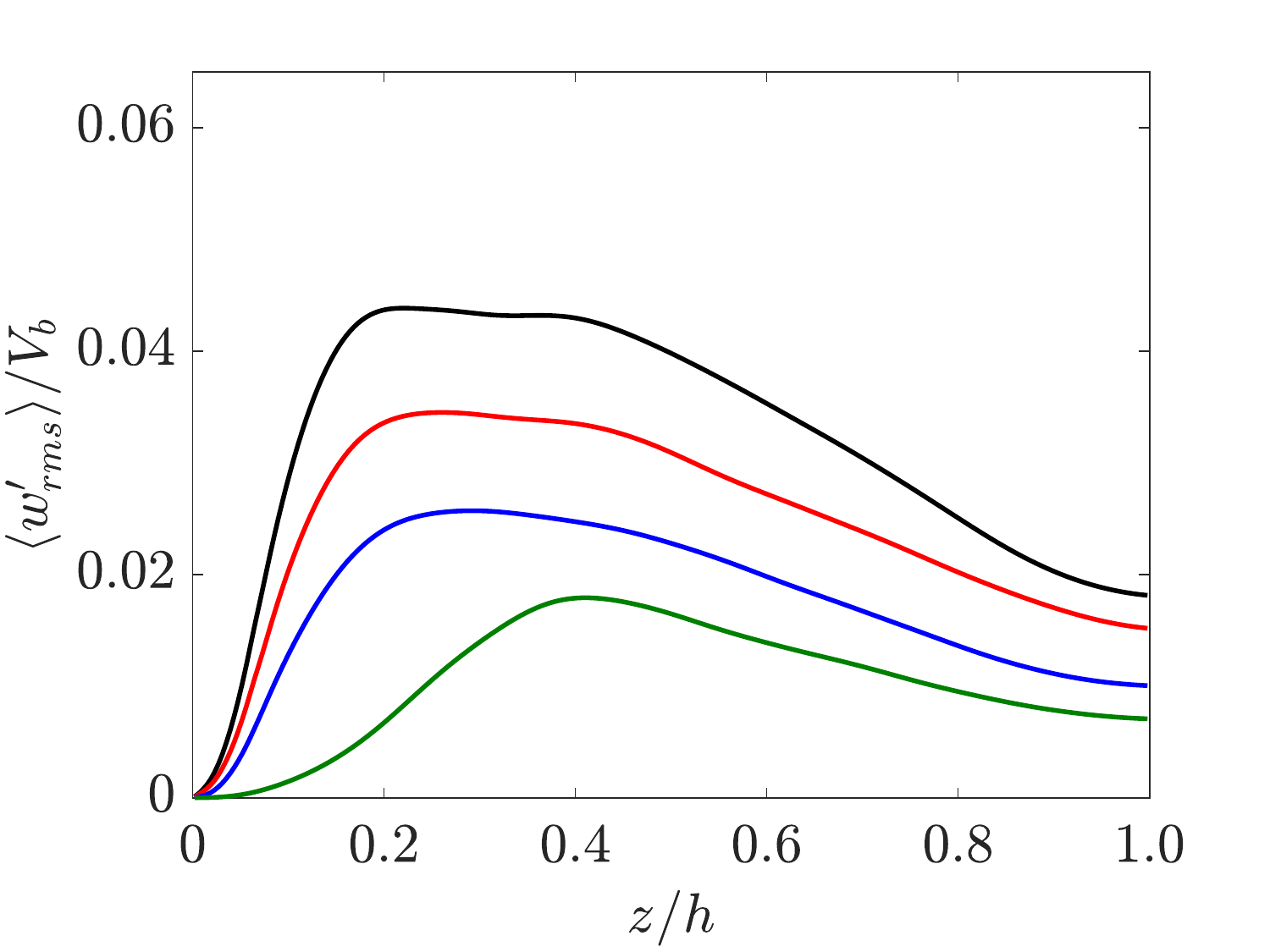} \label{fig:w_rms:b}}
\caption{(a) Stream-wise and (b) wall-normal root-mean-square velocity profiles of a suspension at $\Rey = 3.75$ and $\Ca= 0.009$. Data for different $\Phi$ is plotted with the same color scheme of the previous figures: $\Phi = 0.15$ (black), $0.1$ (red), $0.05$ (blue)  and $0.025$ (green).}
\label{fig:velocity_fluctuations}
\end{figure}

Stream-wise and wall-normal root-mean-square velocity profiles are plotted in figure~\ref{fig:velocity_fluctuations} for different volume fractions at a fixed capillary number $\Ca=0.009$. It can be observed that the former have higher intensities and present some peaks in correspondence with the particle layers, while the latter are less intense and smoother. Both the fluctuation components increase with the particle volume fractions, slightly increase with $\Rey$ and are only marginally affected by $\Ca$ (not shown here).


\begin{figure}[h!]
\centering
\includegraphics[width=0.6\textwidth]{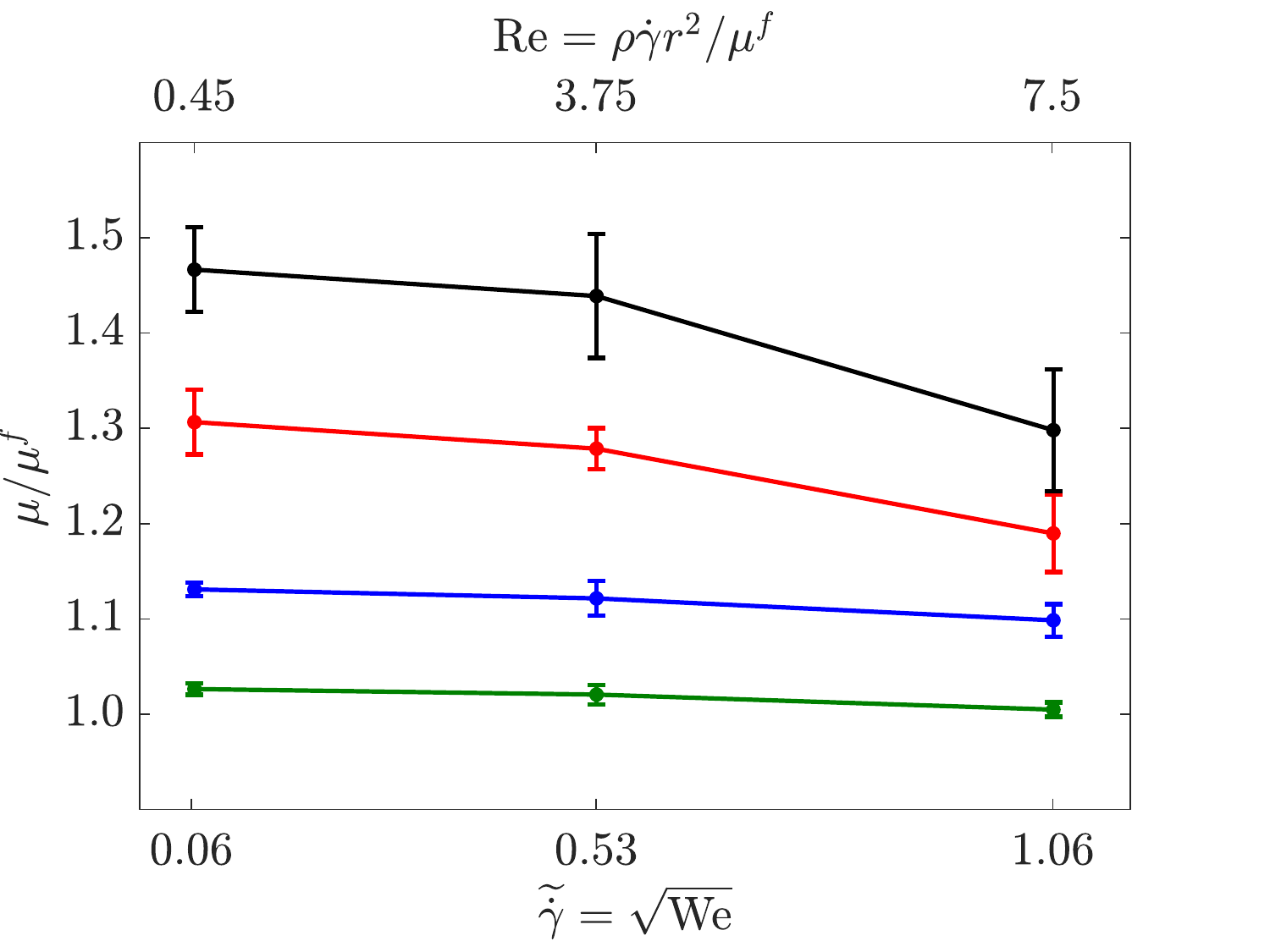}
\caption{Suspension effective viscosity $\mu / \mu^{f}$ as a function of the shear rate. The governing parameters are chosen so that the $\Rey/\Ca$ ratio is constant. The lower x-axis shows the values of the non-dimensional shear rate $\widetilde{\dot{\gamma}}$, while the upper x-axis displays the corresponding particle Reynolds numbers. The black, red, blue and green lines represent data at different solid volume fraction, $\Phi = 0.15$, $0.1$, $0.05$ and $0.025$, respectively, and the error bars are computed based on the stationary distributions of the effective viscosity for the various cases.}
\label{fig:shear_thinning}
\end{figure}

In order to investigate the rheology of the suspension, in figure~\ref{fig:shear_thinning} we select the simulation parameters to mimic experiments in a fixed geometry and same particle material, while changing the bulk shear rate $\dot{\gamma}=3V_b/2h$ (i.e.\ varying the flow rate in the channel). In terms of dimensionless parameters, this corresponds to consider cases at constant $\Rey/\Ca$ ratio  and varying the non-dimensional shear rate $\widetilde{\dot{\gamma}}=\sqrt{\We}=\sqrt{\Rey \, \Ca}$. 
The parameter $\widetilde{\dot{\gamma}}$ can be thought of as the ratio between the convective time scale in the flow and the material deformation time scale.
In the figure we show the suspension relative viscosity $\mu / \mu^{f}$, computed as a cross-sectional average, versus the non-dimensional shear rate $\widetilde{\dot{\gamma}}$. The relative viscosity proves to be a decreasing function of the shear rate, which confirms that the suspension has a shear-thinning behavior, as expected for suspensions of this type of deformable particles. The denser the suspension, the more intense this behavior.


\begin{figure}[h!]
\subfigure[]{\includegraphics[width=0.5\textwidth]{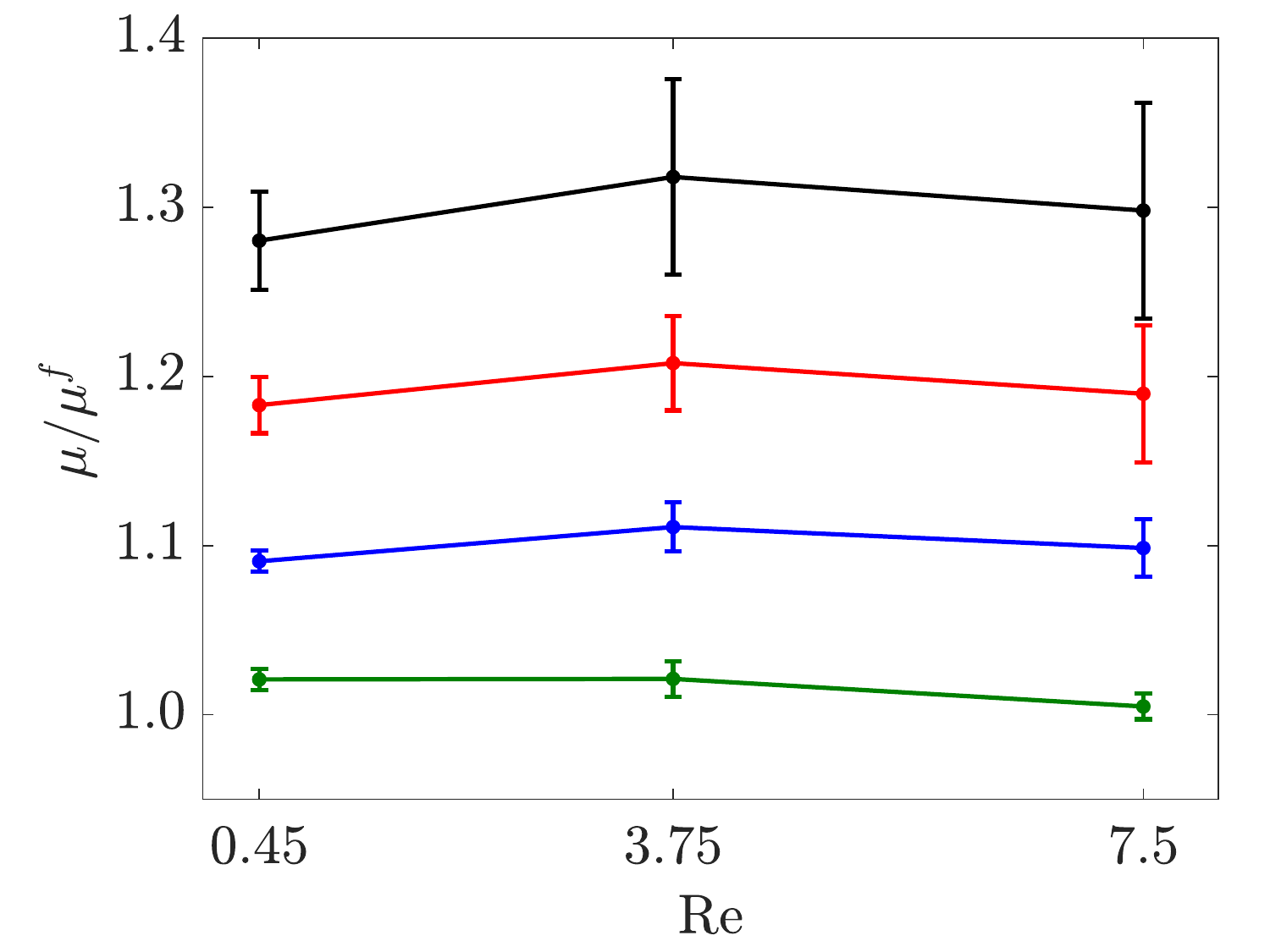} \label{fig:mu_ReCe:a}}
\subfigure[]{\includegraphics[width=0.5\textwidth]{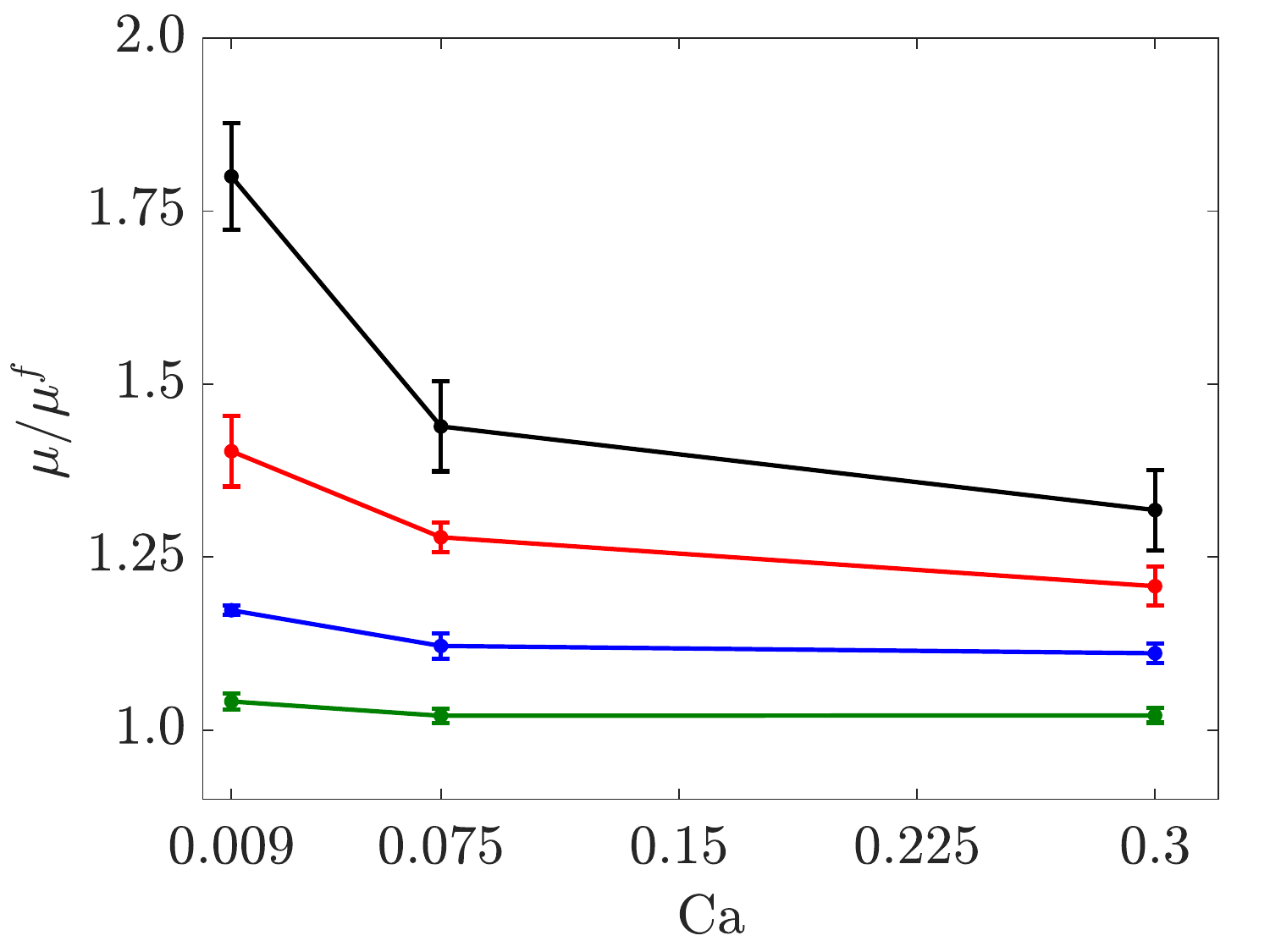} \label{fig:mu_ReCe:b}}
\caption{Effective viscosity $\mu / \mu^{f}$ as a function of (left) the particle Reynolds number $\Rey$, at fixed $\Ca = 0.3$, and (right) the capillary number $\Ca$, at fixed $\Rey = 3.75$. The plots illustrate data for different total volume fractions, $\Phi = 0.15$ (black), $0.1$ (red), $0.05$ (blue) and $0.025$ (green).}
\label{fig:mu_ReCa}
\end{figure}

According to the definition of the Weber number $\We = \Rey \, \Ca$, both the capillary and Reynolds numbers may contribute to this shear-thinning. We analyze therefore the dependency of the suspension relative viscosity on these two parameters by fixing one and varying the other. In figure~\ref{fig:mu_ReCa} we plot the relative viscosity versus $\Rey$ at constant $\Ca=0.3$ in the left panel, and versus $\Ca$ at constant $\Rey=3.75$ in the right panel, for different volume fractions $\Phi$. The particle Reynolds number varies between 0.45 and 7.5, while the capillary number between 0.009 and 0.3. The data indicate that the relative viscosity weakly depends on $\Rey$, while it appears a decreasing function of $\Ca$. Hence, for the range of parameters investigated in this study, the viscosity depends more on the capillary than on the Reynolds number.


\begin{figure}[h!]
\subfigure[]{\includegraphics[width=0.5\textwidth]{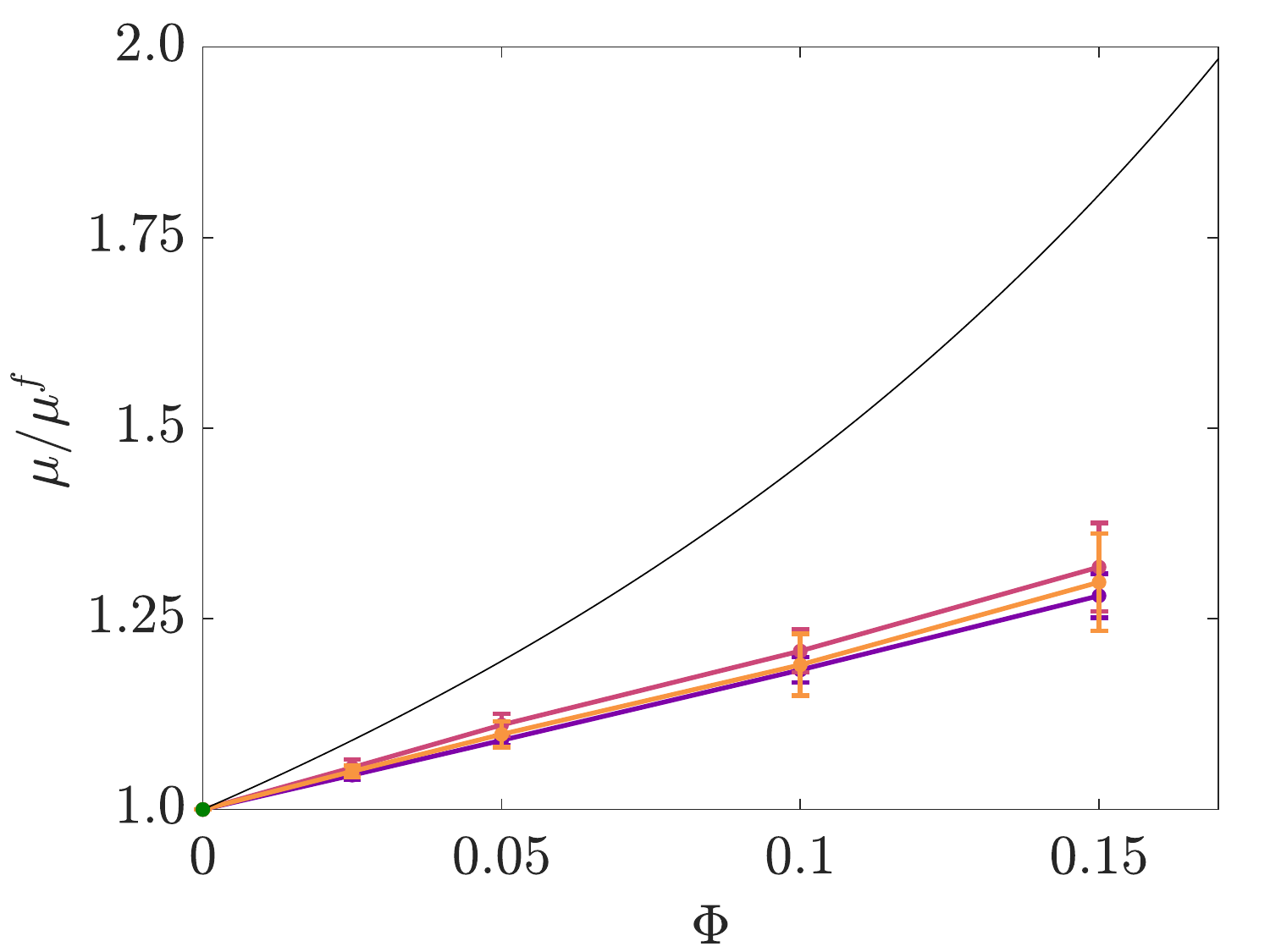} \label{fig:mu_Phi:a}}
\subfigure[]{\includegraphics[width=0.5\textwidth]{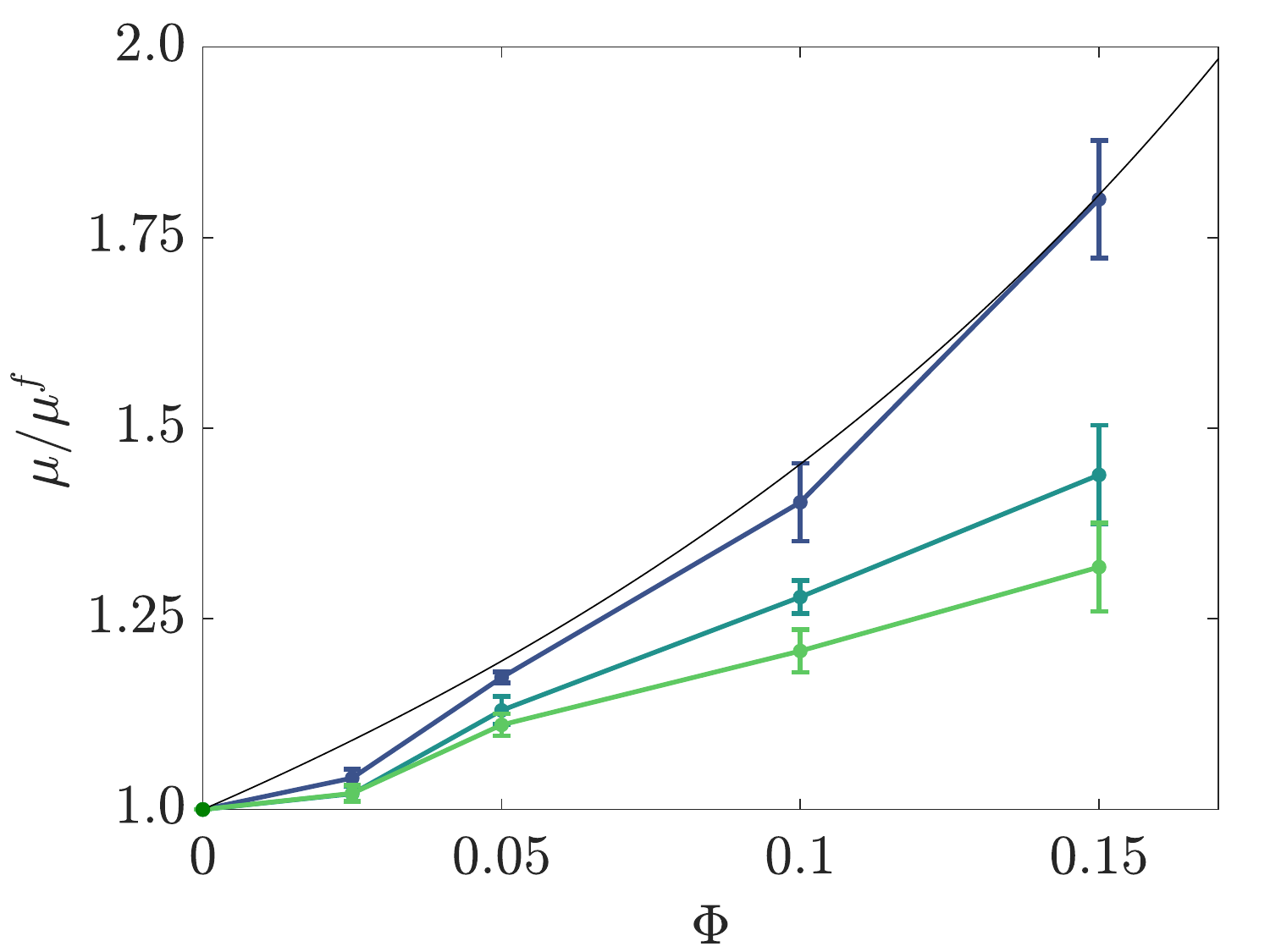} \label{fig:mu_Phi:b}}
\caption{Effective viscosity versus suspension volume fraction. The panels depict the same data of figure~\ref{fig:mu_ReCa}, this time plotted using the volume fraction as horizontal coordinate. Cases for (a) different Reynolds and (b) different capillary numbers are shown. In the left panel, violet, orange and pink represent simulations at $\Rey = 0.45$, $3.75$ and $7.5$, respectively. Likewise, in the right panel each color represents a different capillary number, $\Ca = 0.009$ (blue), $0.075$ (dark green) and $0.3$ (light green). Finally, the black line in the figures represents the viscosity of rigid particle suspensions predicted by the Eilers fit.}
\label{fig:mu_Phi}
\end{figure}

To gain further insight into the suspension behavior, we plot the same data as explicit functions of the volume fractions in figure~\ref{fig:mu_Phi}. It can be noted that cases at different Reynolds numbers (left panel) cluster together and do not show any significant difference, while the effective viscosity curves vary as $\Ca$ changes (right panel). In particular, the effective viscosity decreases as $\Ca$ increases and this effect is more pronounced at high $\Phi$.
In figure~\ref{fig:mu_Phi} we also plot the Eilers fit, which is valid for rigid spherical particles in the inertialess limit, both for dense and dilute suspensions:
\begin{equation}
\frac{\mu}{\mu^{f}} = \Big[ 1 + B_E \frac{\Phi}{1 - \Phi / \Phi_{m}} \Big]^2.
\end{equation}
In the previous equation $\Phi_m = 0.58 \sim 0.63$ is the geometrical maximum packing, and $B_E = 1.25 \sim 1.7$ a fitting coefficient. $\Phi_m = 0.58$ and $B_E = 1.7$ appear to be good values for our simulations. It can be remarked in figure~\ref{fig:mu_Phi:b} that the most rigid case essentially follows the Eilers fit, especially at high volume fractions. However, interestingly for small volume fraction, e.g. $\Phi=0.025$, the Eilers fit does not provide a good approximation and the suspension viscosity is closer to that of the unladen case. We explain this peculiar behavior by the shear-induced migration of deformable particles. In particular, deformable particles tend to migrate towards the middle of the channel, where the shear rate vanishes. In other words, to avoid deformation, particles tend to avoid the high shear rate region near the wall and stay near the centerline. This is a direct consequence of what previously observed in figure~\ref{fig:Phi_over_z}, further showing that simple correlations extracted from a Couette geometry may be not enough to fully describe more complex flow configurations such as the present one. 


\begin{figure}[h!]
\subfigure[]{\includegraphics[width=0.5\textwidth]{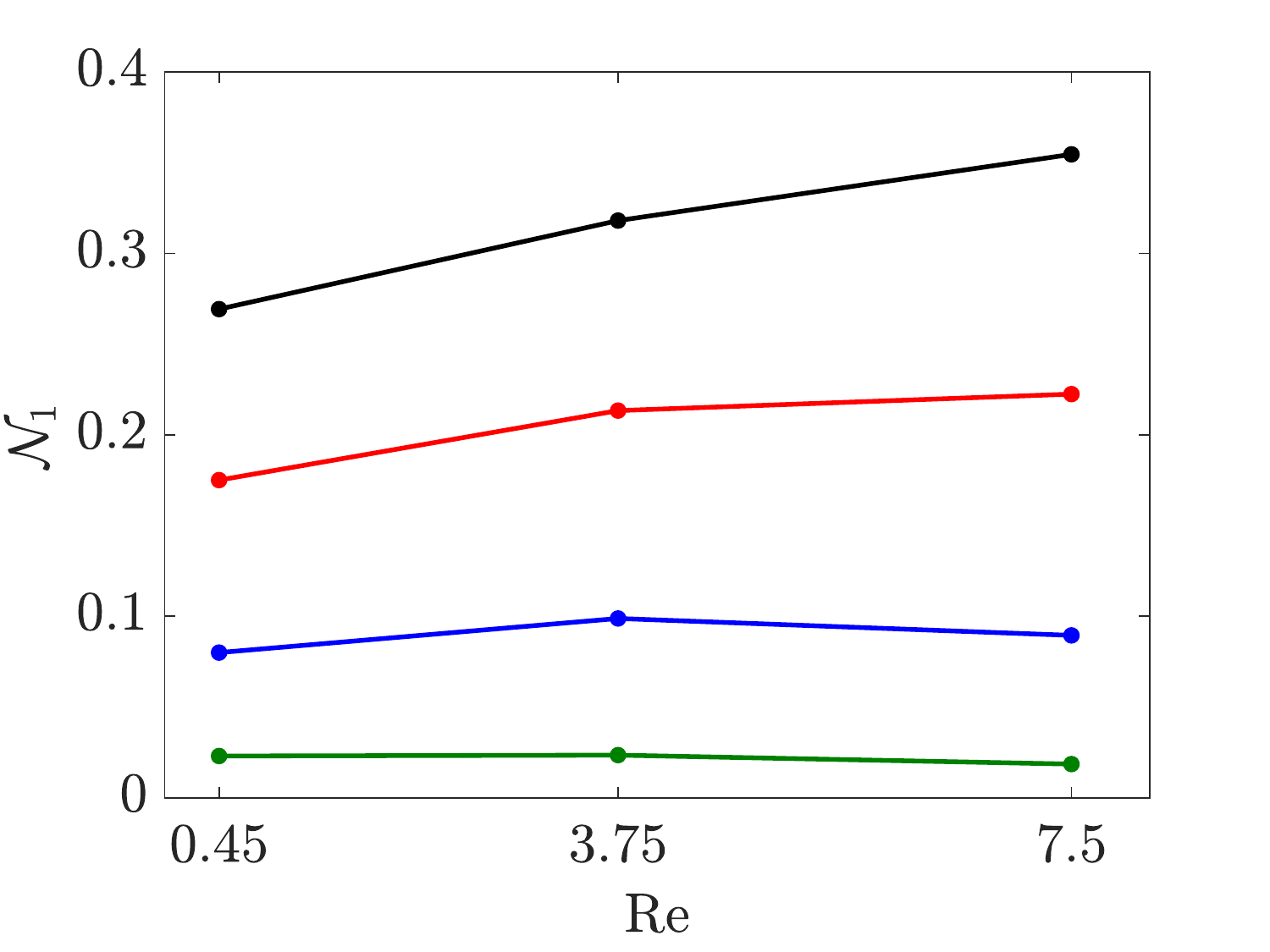} \label{fig:N1_Re_Phi}}
\subfigure[]{\includegraphics[width=0.5\textwidth]{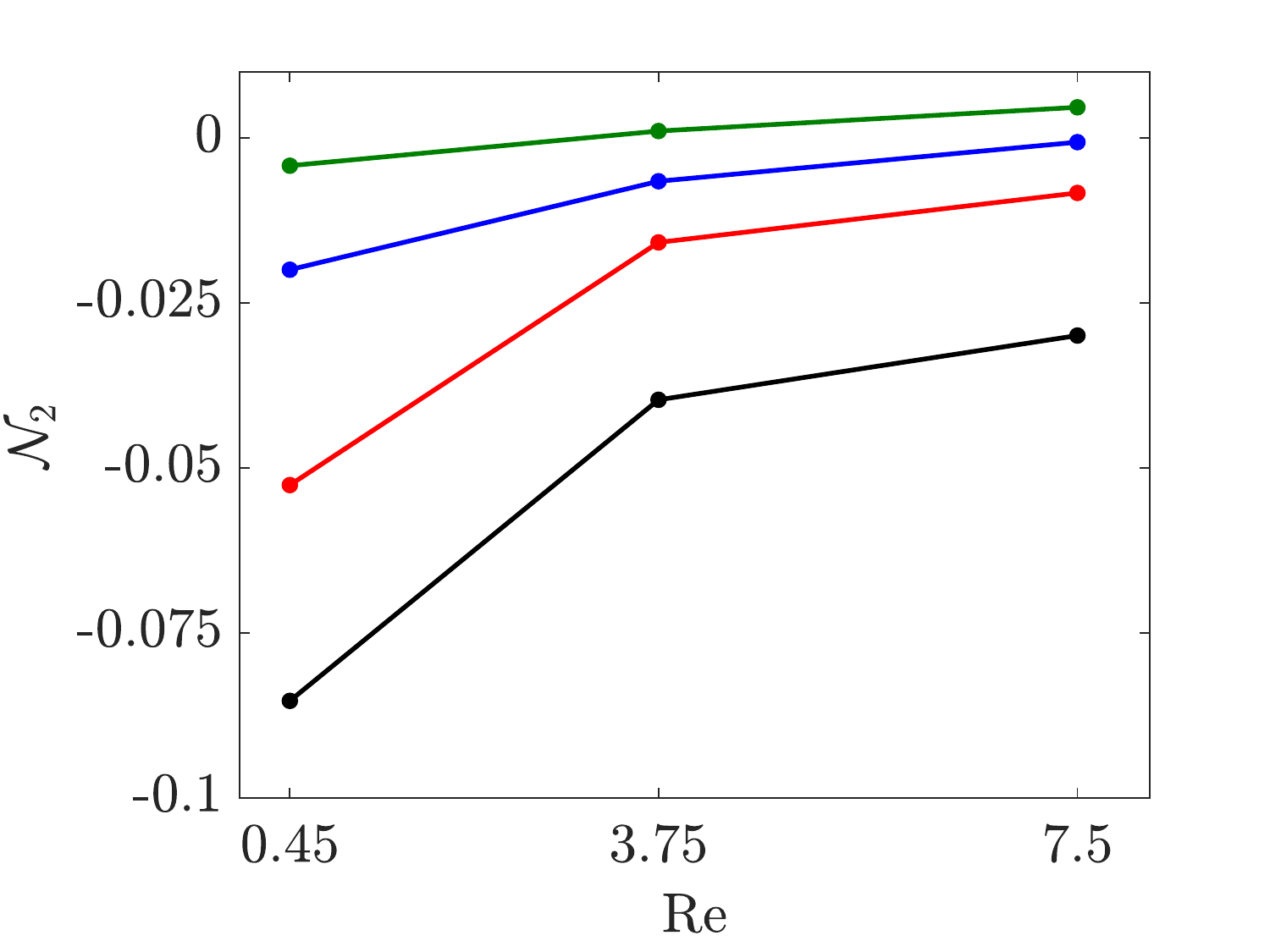} \label{fig:N2_Re_Phi}}
\caption{(a) First and (b) second normal stress differences of a solution at $\Ca=0.3$, plotted versus $\Rey$. The different colors represent cases at different $\Phi$, with the usual color scheme.}
\label{fig:N1_N2_Re_Phi}
\end{figure}

\begin{figure}[h!]
\subfigure[]{\includegraphics[width=0.5\textwidth]{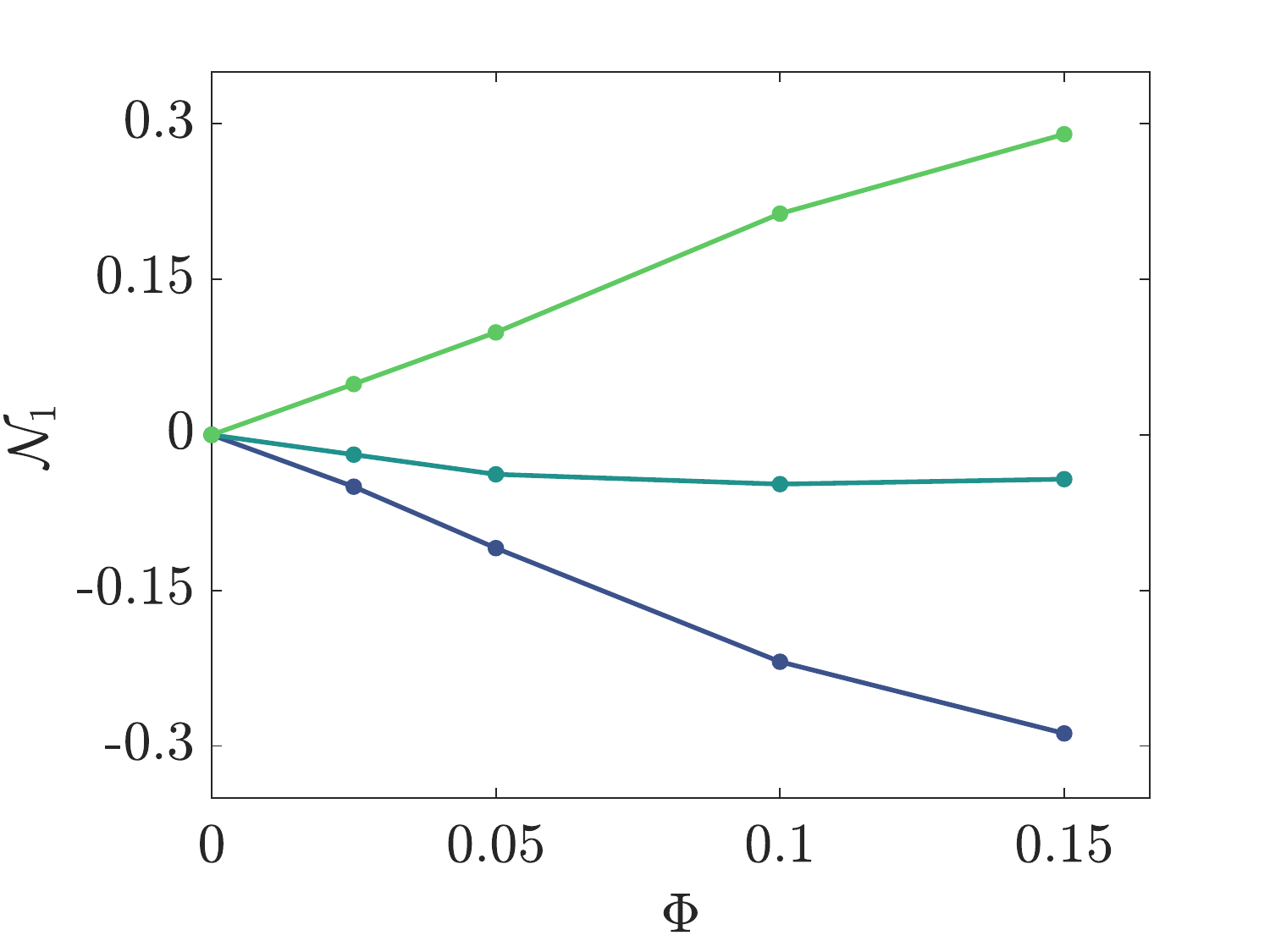} \label{fig:N1_Phi_Ca}}
\subfigure[]{\includegraphics[width=0.5\textwidth]{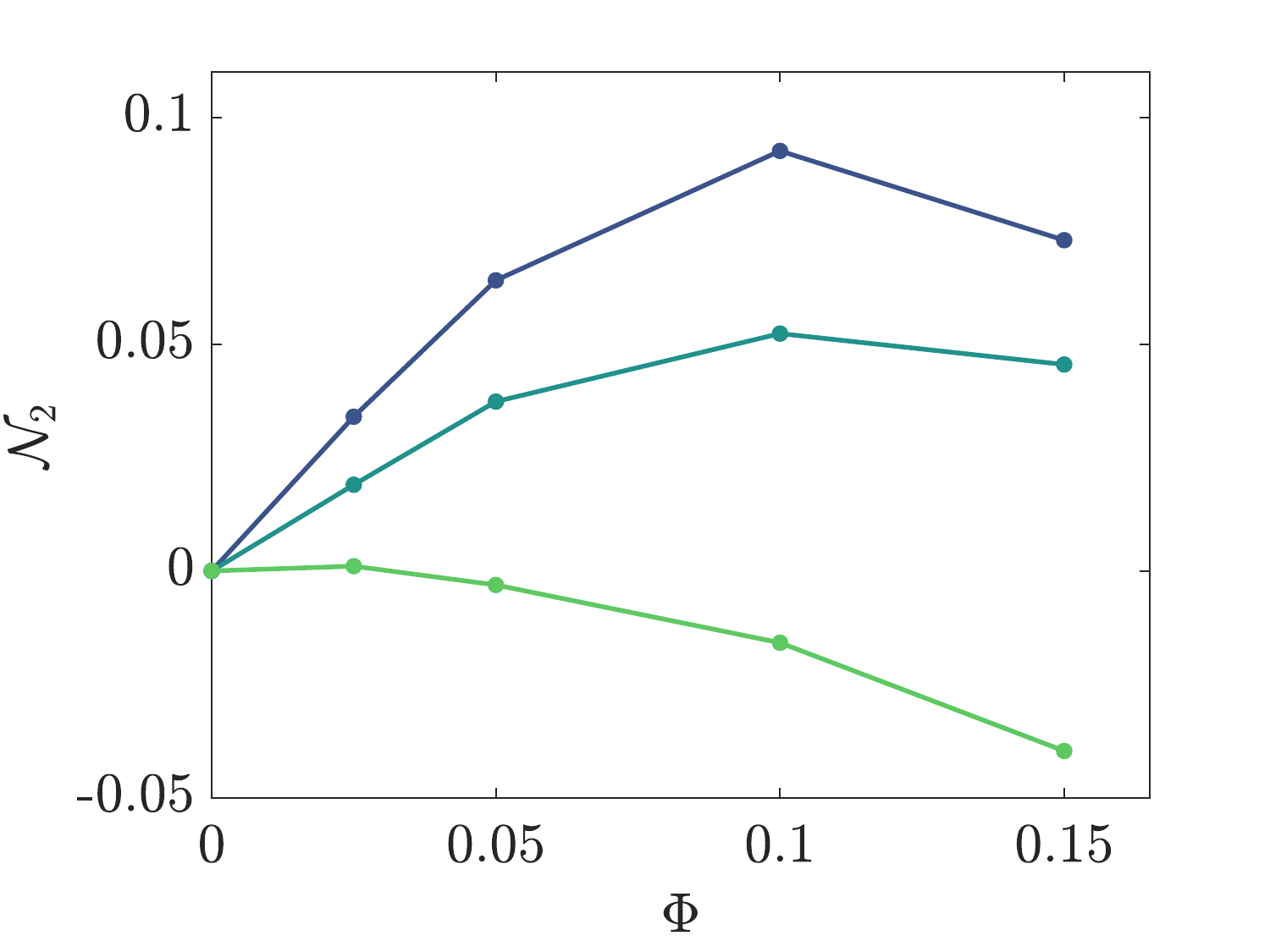} \label{fig:N2_Phi_Ca}}
\caption{(a) First and (b) second normal stress differences of a solution at $\Rey=3.75$, plotted as functions of $\Phi$. The different colors represent data at $\Ca = 0.009$ (blue), $0.075$ (dark green) and $0.3$ (light green).}
\label{fig:N1_N2_Phi_Ca}
\end{figure}

The elastic nature of the particles is evaluated here in terms of the first and second normal stress differences, defined as:
\begin{eqnarray}
\label{eq:N1_N2}
\mathcal{N}_1 &=& \langle \langle \sigma_{22} - \sigma_{33}  \rangle \rangle = \langle \langle \phi G B_{22} + 2 \mu^f \frac{\partial v}{\partial y} - \phi G B_{33} - 2 \mu^f \frac{\partial w}{\partial z} \rangle \rangle, \\
\mathcal{N}_2 &=& \langle \langle \sigma_{33} - \sigma_{11}  \rangle \rangle = \langle \langle \phi G B_{33} + 2 \mu^f \frac{\partial w}{\partial z} - \phi G B_{11} - 2 \mu^f \frac{\partial u}{\partial x} \rangle \rangle.
\end{eqnarray}
These two quantities are a measure of the anisotropy of the flow and are important when characterizing the non-Newtonian behaviour of fluids, since together with the effective viscosity can provide information of the full stress tensor. Furthermore, when considering particle suspensions, they are usually associated to the particle migration.
$\mathcal{N}_1$ and $\mathcal{N}_2$ are reported in figure~\ref{fig:N1_N2_Re_Phi} as a function of the Reynolds number for different volume fractions at the largest value of $\Ca$ considered, and in figure~\ref{fig:N1_N2_Phi_Ca} as a function of the volume fraction for different values of $\Ca$. When the capillary number is fixed and sufficiently large, $\mathcal{N}_1$ is positive, $\mathcal{N}_2$ negative, and their ratio $-\mathcal{N}_1/\mathcal{N}_2$ larger than unity. This is consistent with what found for other elastic fluids, such as polymeric solution \cite{shahmardi_zade_ardekani_poole_lundell_rosti_brandt_2019a}, capsule \cite{matsunaga_imai_yamaguchi_ishikawa_2016a} and fiber suspensions \cite{banaei_rosti_brandt_2020a}. We observe that, $\mathcal{N}_1$ grows as the Reynolds number increases, while $\mathcal{N}_2$ reduces, thus enhancing their difference. Very different results are observed when the capillary number is varied. Indeed, when $\Ca$ is small (almost rigid particles) $\mathcal{N}_1$ is actually negative and becomes positive only for $\Ca \approx 0.1$. This is consistent with what typically found when considering rigid particle suspensions \cite{kulkarni_morris_2008a}. Also, $\mathcal{N}_2$ assumes an opposite sign when the capillary number is small, in the limit of rigid particles. In general, a negative value for $\mathcal{N}_1$ can be interpreted as a signature of the importance of particle interactions and contacts, which is dominant for the rigid ones but negligible for the deformable ones, when $\mathcal{N}_1$ even changes sign. Also, the increase of the first normal stress difference with $\Rey$ and $\Phi$ for the deformable cases indicates a stronger tendency of the fluid to displace the particles towards the centerline ~\cite{banaei_rosti_brandt_2020a}.

\begin{figure}[h!]
\subfigure[]{\includegraphics[width=0.33\textwidth]{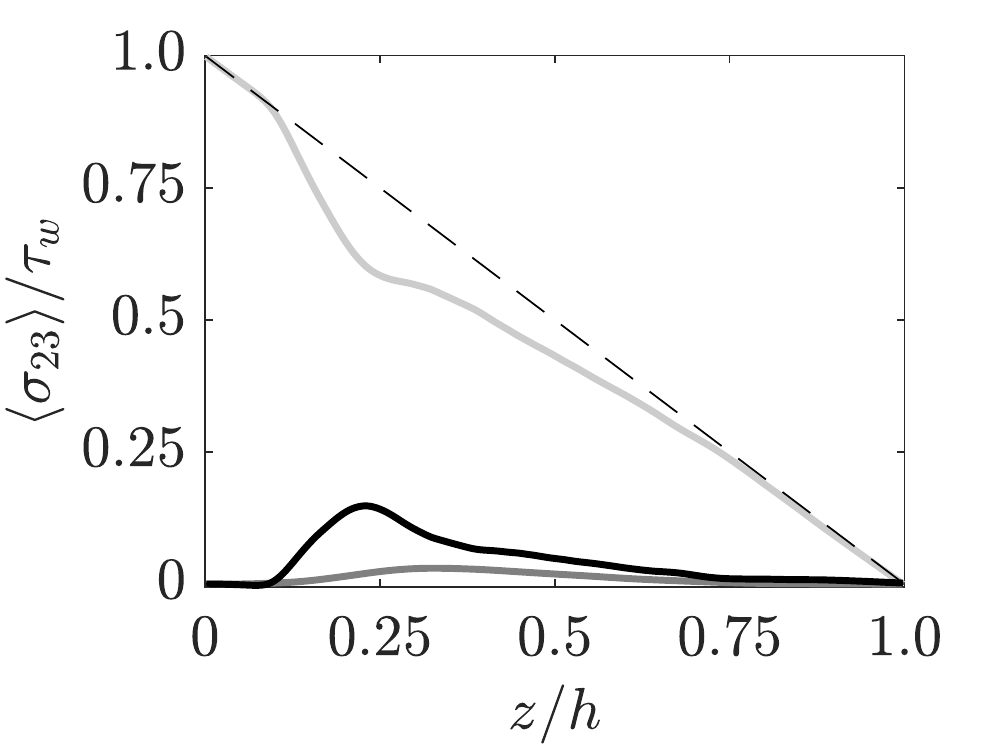}\label{fig:stress_z:a}}
\subfigure[]{\includegraphics[width=0.33\textwidth]{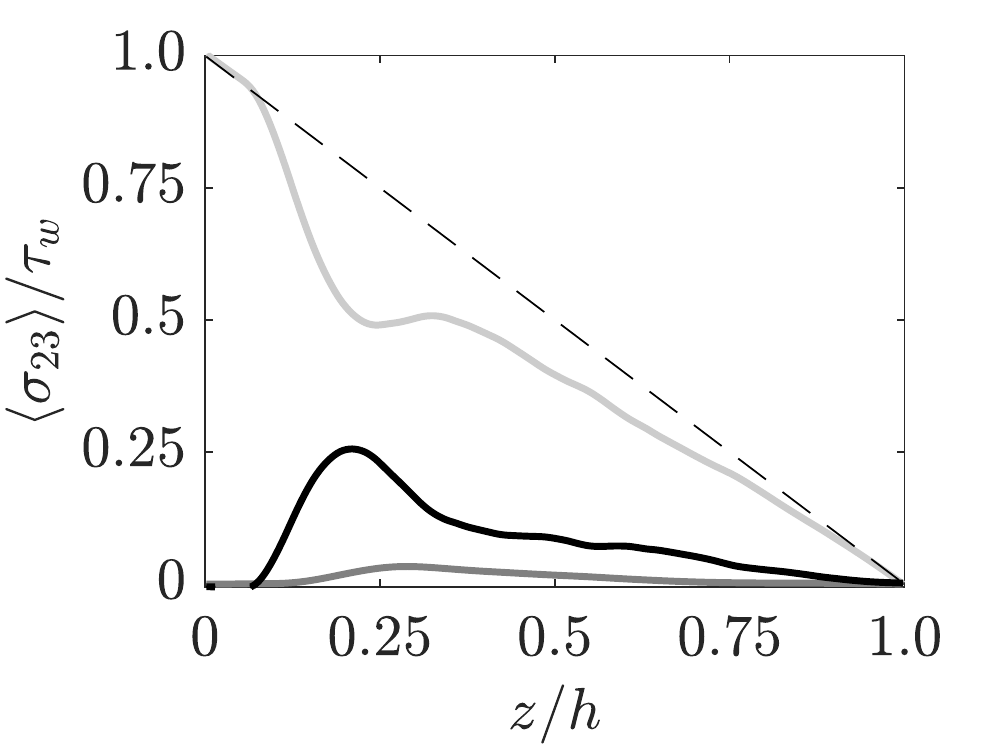}\label{fig:stress_z:b}}
\subfigure[]{\includegraphics[width=0.33\textwidth]{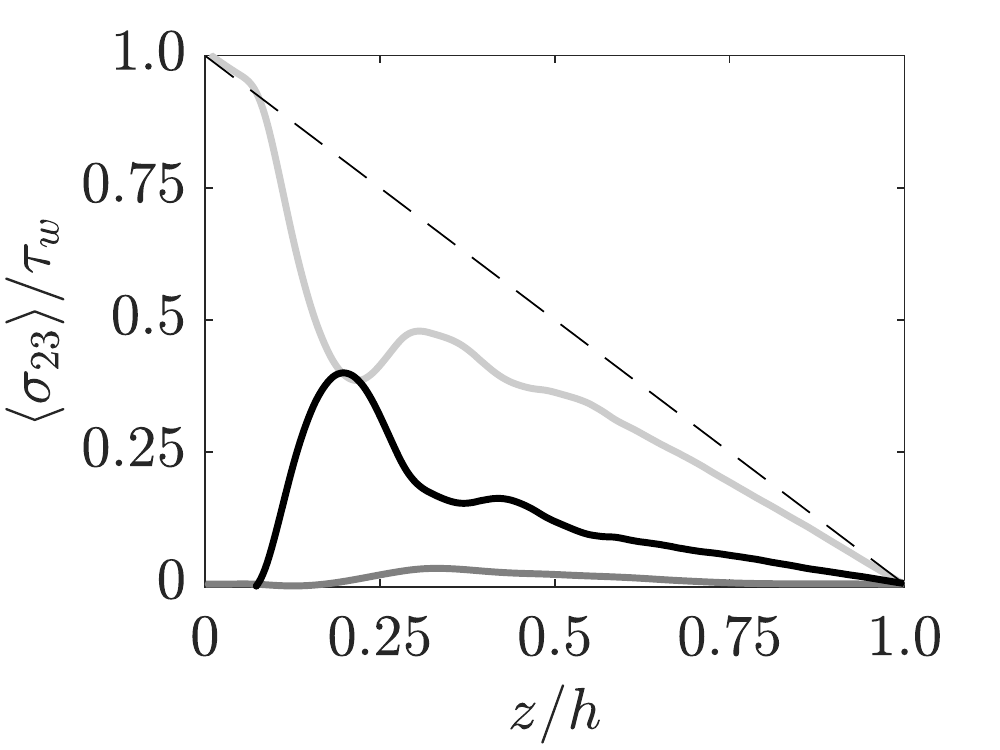}\label{fig:stress_z:c}}
\caption{Decomposition of the total shear stress $\langle \sigma_{23} \rangle$ in its contributions, normalized by the total wall value $\tau_w = \langle \sigma_{23} \rangle|_{w}$. The data refer to a suspension at $\Rey = 3.75$, $\Ca = 0.3$ and (a) ${\Phi} = 0.05$, (b) $0.1$ and (c) $0.15$. In the plots, the light gray, dark gray and black solid lines represent the viscous, Reynolds and particle hyperelastic stresses respectively, while the black dashed line represents the total shear stress.}
\label{fig:stress_z}
\end{figure}

The increase of the relative viscosity is linked to an increase in the stress across the channel. The total shear stress $\sigma_{23}$ can be decomposed into the sum of the fluid $\sigma_{23}^{f}$ and particle stress $\sigma_{23}^{s}$, as in equation~\ref{eq:phi-stress}. The mean fluid stress is the sum of the viscous and Reynolds stresses, while the solid one is the sum of the viscous, Reynolds, hyperelastic and collision stresses \cite{rosti_brandt_2018a, rosti_de-vita_brandt_2019a, picano2015turbulent}
\begin{equation}
\bra{\sigma_{23}} = \underbrace{\bra{ \left( 1 - \phi \right) \left( \mu^f \frac{\partial v}{\partial z} - \rho v'w' \right)}}_{\rm fluid~stress} + \underbrace{ \bra{\phi \left( \mu^f \frac{\partial v}{\partial z} - \rho v'w' + G B_{23} \right) + \int_{0}^{z} f_{y} dz}}_{\rm solid~stress}.
\end{equation}
If we consider the monolithic formulation independent of the actual phase, the total shear stress can also be decomposed into the sum of the total viscous, total Reynolds, hyperelastic and collision stresses in the following way
\begin{equation}
\bra{\sigma_{23}} = \underbrace{\bra{\mu^f \frac{\partial v}{\partial z}}}_{\rm viscous~stress} + \underbrace{\bra{ - \rho v'w'}}_{\rm Reynolds~stress} + \underbrace{\bra{G B_{23}}}_{\rm elastic~stress} + \underbrace{\bra{\int_{0}^{z} f_{y} dz}}_{\rm collision~stress}.
\end{equation}
In particular, the total shear stress $\bra{\sigma_{23}}$ in a plane channel flow is a linear function, equal to the total wall shear stress at the wall and null in the centerline. All the shear stress contributions are averaged and displayed in figure~\ref{fig:stress_z} as functions of the wall-normal distance $z$ for the cases at $\Rey = 3.75$, $\Ca=0.3$ and volume fractions $\Phi=0.05$, $0.1$ and $0.15$. The stresses are normalized by the total wall value. In the figures, the light gray, dark gray and black color represent the viscous, Reynolds, and particle stresses respectively. The collision stress contribution is negligible in all the cases considered here, thus not shown in the plots for greater clarity.

As expected, the viscous stress is the only not vanishing component at the wall. For all cases, since we are dealing with laminar suspension flows, the viscous stress is the dominant contribution. It however decreases in correspondence of the location of the maximum particle concentrations (see figure~\ref{fig:Phi_over_z}), where the particle stress reaches its maximum. As the volume fraction is increased, the viscous stress becomes smaller, which is compensated by an increase in the particle stress contribution, which eventually accounts for up to $50\%$ of the total shear stress in the central region of the channel at $\Phi = 0.15$. Concerning the Reynolds stresses, it can be observed that they are almost negligible in every case shown here, rising slightly as $\Phi$ and $\Ca$ are increased, with a maximum located at around $z = 0.3 \ h$.


\begin{figure}[h!]
\centering
\subfigure[]{\includegraphics[width=0.8\textwidth]{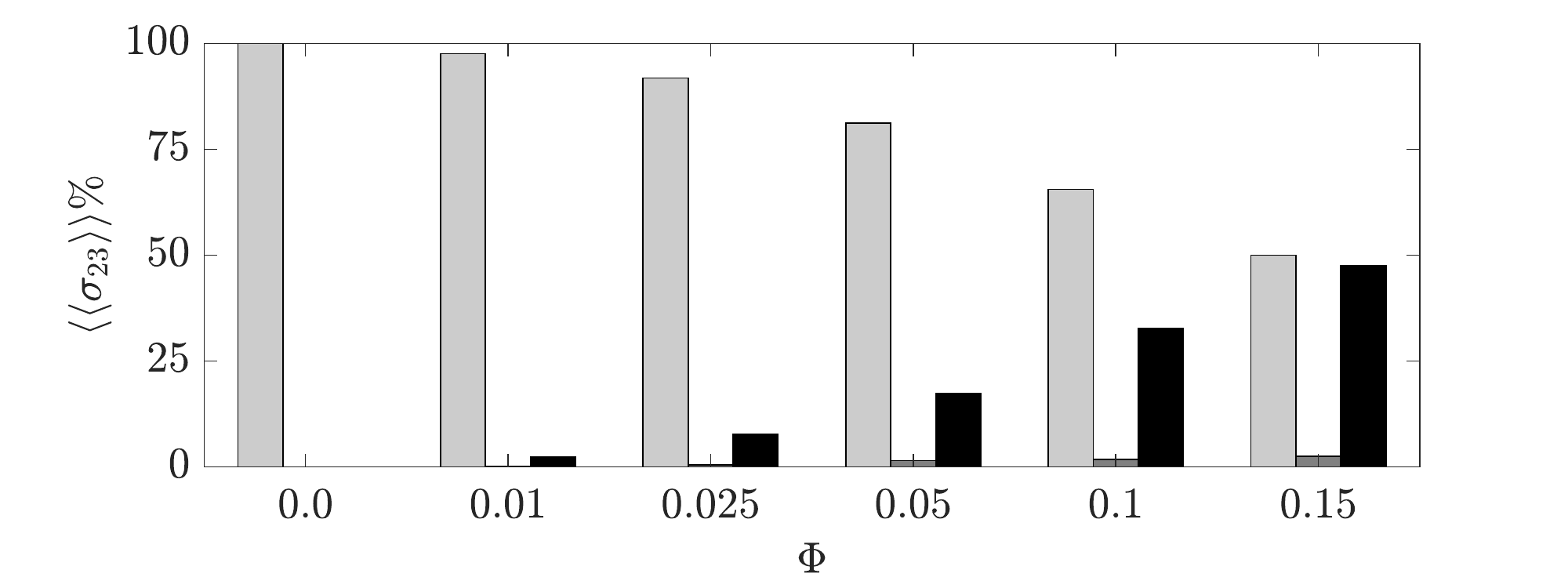} \label{fig:budget_Phi:a}}
\subfigure[]{\includegraphics[width=0.8\textwidth]{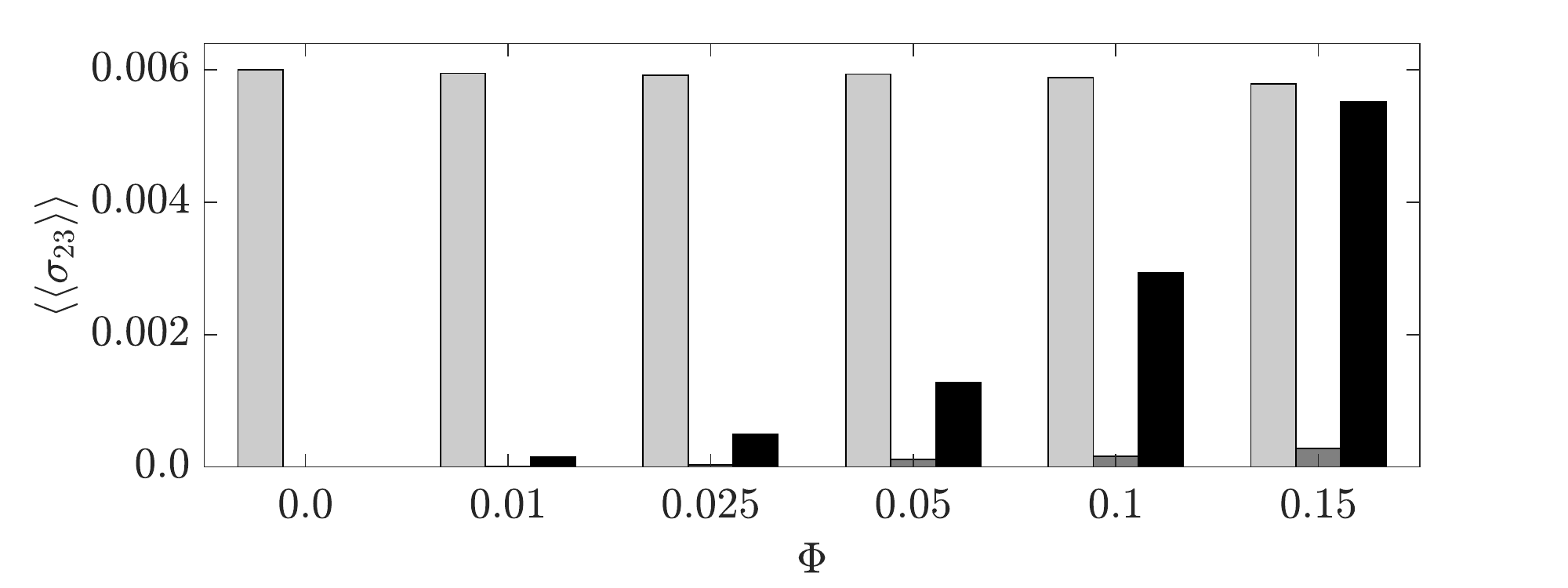} \label{fig:budget_Phi:b}}
\caption{Total volume-averaged shear stress $\langle \langle \sigma_{23} \rangle \rangle$ decomposed into viscous (light gray), Reynolds (dark gray) and particle hyperelastic (black) contribution. The histograms give information about a suspension at fixed $\Rey = 3.75$ and $\Ca = 0.009$, for different volume fractions, both in (a) percentages and (b) absolute values.}
\label{fig:budget_Phi}
\end{figure}

\begin{figure}[h!]
\subfigure[]{\includegraphics[width=0.5\textwidth]{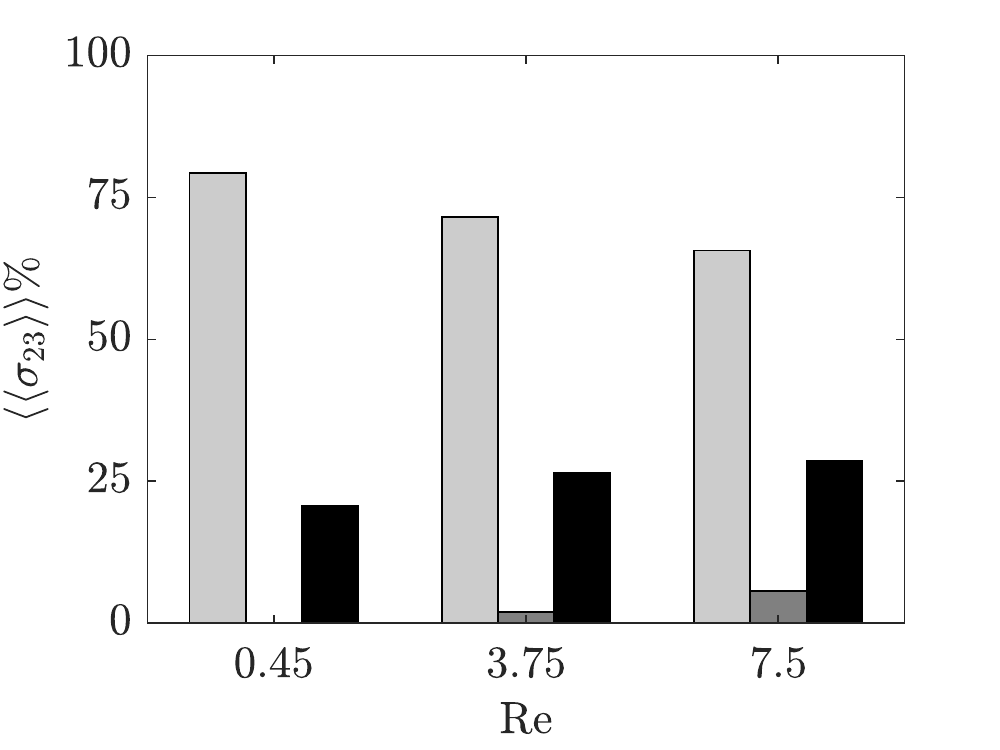} \label{fig:budget_ReCa:a}}
\subfigure[]{\includegraphics[width=0.5\textwidth]{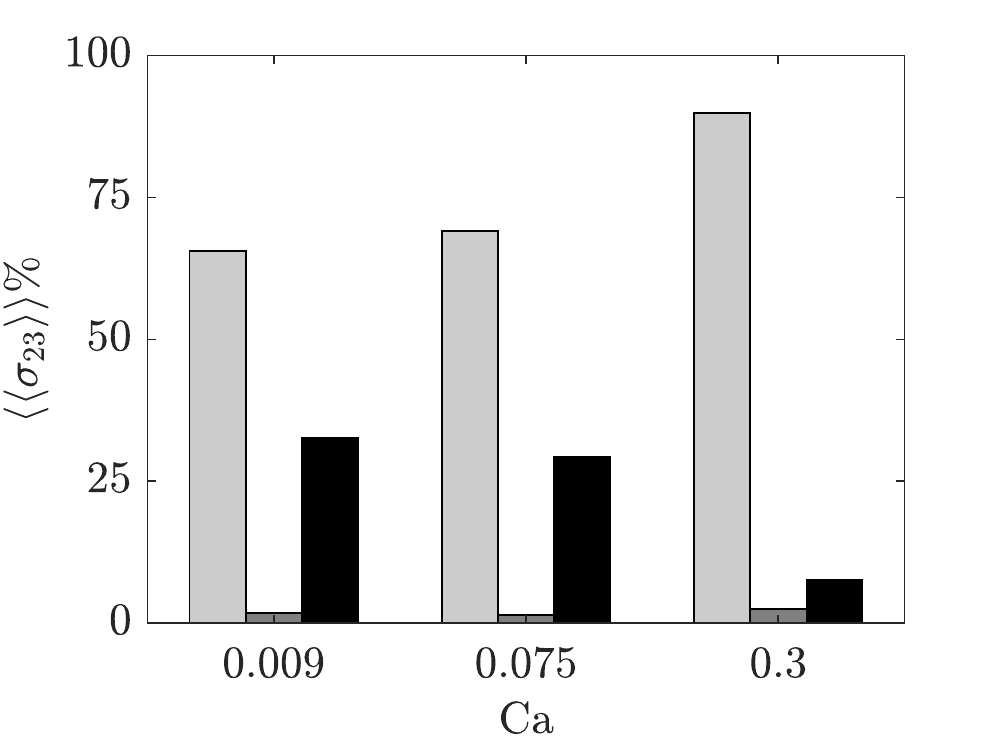} \label{fig:budget_ReCa:b}}\\
\subfigure[]{\includegraphics[width=0.5\textwidth]{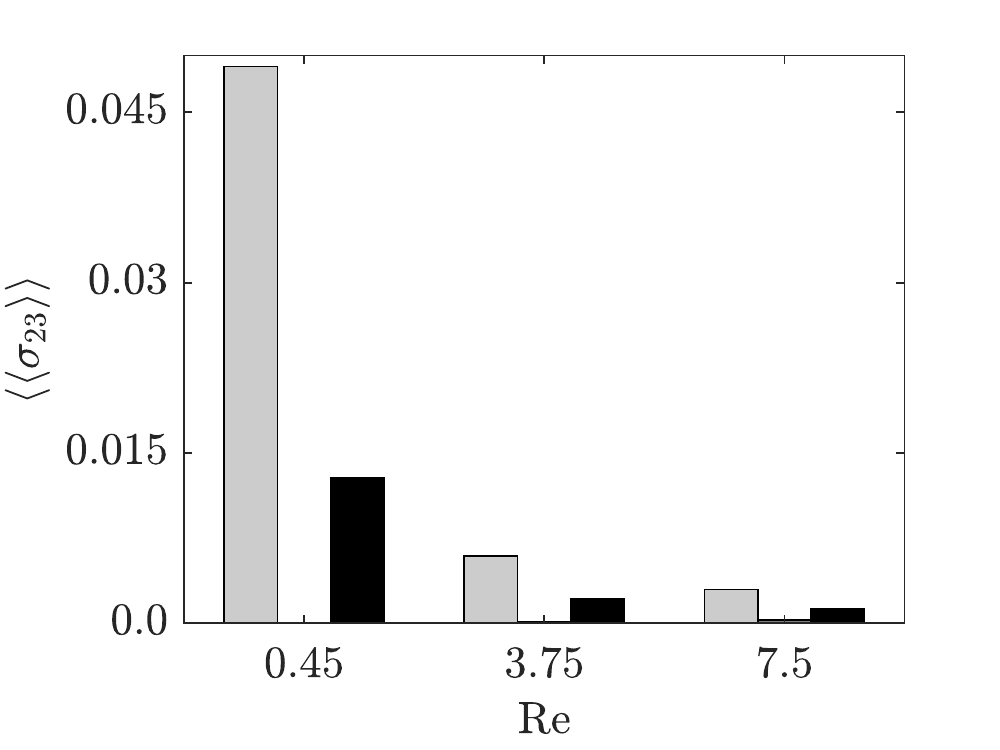} \label{fig:budget_ReCa:c}}
\subfigure[]{\includegraphics[width=0.5\textwidth]{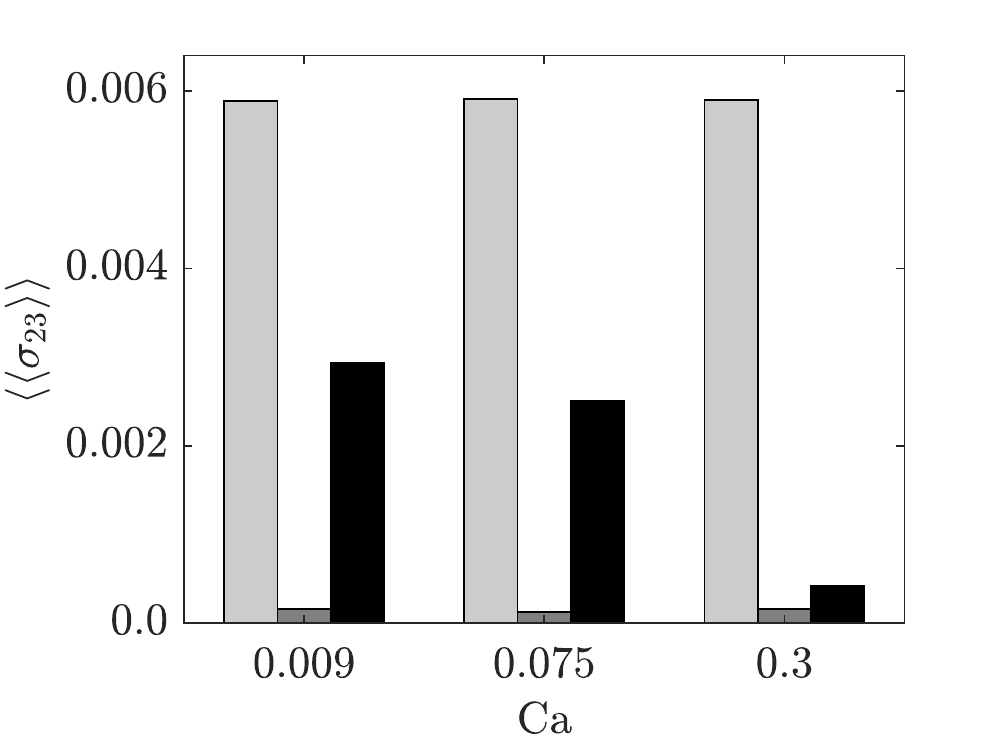} \label{fig:budget_ReCa:d}}
\caption{Contributions to the total volume-averaged shear stress $\langle \langle \sigma_{23} \rangle \rangle$ plotted versus $\Rey$ and $\Ca$. The color scheme is the same as in the previous figures, with light gray, dark gray and black bars indicating the viscous, Reynolds, and particle hyperelastic contribution respectively. Panels (a) and (c) display data at $\Ca = 0.3$ and $\Phi = 0.1$ for different Reynolds numbers, both in relative and absolute terms. Panels (b) and (d) illustrate the same quantities for a suspension at $\Rey = 3.75$, $\Phi = 0.1$ and different capillary numbers.}
\label{fig:budget_ReCa}
\end{figure}

In figure~\ref{fig:budget_Phi} we summarize the stress budgets for fixed $\Rey=3.75$ and $\Ca=0.009$ by showing the volume-averaged contributions of all the components of the total shear stress. Viscous (light gray), Reynolds (dark gray), and particle stresses (black) are plotted both as percentages of the total stress and as bulk values. Comparing the budgets at different volume fractions, we observe that as the volume fraction increases, the bulk value of the viscous stress remains essentially the same, while the particle-induced stress increases, becoming of the same order of magnitude as the viscous stress at $\Phi=0.15$. The Reynolds shear stress, although slightly increasing, remains secondary.

Finally, we display in figure~\ref{fig:budget_ReCa} the budgets for different $\Rey$ at fixed $\Ca$ and vice versa. At fixed $\Ca$, when increasing $\Rey$ we note that the particle stress increases in importance with respect to the viscous stress when inertia becomes more relevant. Concerning the increase of $\Ca$ at fixed $\Rey$, the total amount of the viscous stress is essentially unchanged, but the particle stress amount reduces, so that, in relative terms, the viscous stress becomes more dominant.

\section{Conclusion}
\label{sec:Conlcusion}

We studied suspensions of deformable particles to characterize the effect of elasticity in the presence of finite inertia and non-uniform shear rate. The flow is characterized by three non-dimensional parameters: the solid volume fraction, the Reynolds and the capillary numbers. The suspensions proved to have a shear-thinning behavior as the relative viscosity happens to be a decreasing function of the shear rate even for finite Reynolds numbers. The denser the suspension, the more this behavior is emphasized.

We started examining the averaged mean velocity, particle distribution, and velocity fluctuation profiles along the wall-normal coordinate. To begin with, the mean velocity is very close to the parabolic solution of the Navier-Stokes equations but exhibits a blunt shape typical of shear-thinning fluids. It does not vary considerably in the cases under investigation. The particle distribution profiles show some concentration peaks at off-center locations, which point to a layered structure. These profiles depend mostly on the total volume fraction and the capillary number. On the other hand, the velocity fluctuations are more sensitive to changes of $\Phi$ and $\Rey$ rather than $\Ca$.

A separate analysis of the dependency of the suspension relative viscosity on the Reynolds and capillary numbers was carried out by fixing one parameter and varying the other. The data indicate that the relative viscosity weakly depends on the Reynolds number, while it appears a decreasing function of the deformability parameter. Hence, for the range of parameters investigated in this study, the viscosity is much more dependent on the capillary than on the Reynolds number.

We showed that for small volume fraction the Eilers fit does not provide a good approximation of the suspension viscosity, which is closer to that of the unladen case. We explained this peculiar behavior by the fact that deformable particles tend to migrate towards the middle of the channel, where the shear rate is low. In other words, to avoid deformation, particles tend to stay away from the high shear rate region near the wall and cluster towards the centerline of the channel. Because of the small value of the shear rate in the core region, the suspended phase does not contribute to the rise in the effective viscosity as in the rigid case, where no migration occurs. This phenomenon is particularly marked for very dilute suspensions, where a whole region near the wall is empty of particles at stationarity. When the volume fraction is increased, this near-wall region is gradually occupied. This is due to the mutual interactions between particles, increasing in number as the volume fraction increases, forcing them to occupy the whole channel in dense configurations. Moreover, as the capillary number increases, this deformation-driven migration is more noticeable. Indeed, the more deformable particles are more sensitive to shear rate variations, and they can modify their shape more easily to accommodate a greater number of particles in the channel center. This behaviour makes the suspension highly non-uniform across the domain, demonstrating the difference with what usually measured and observed in a more standard Couette rheological configuration.

The analysis of the first and second normal stress differences allowed to better evaluate the elastic properties of the particles, and their dependency on the Reynolds and capillary numbers was investigated. We showed that when the capillary number is fixed and sufficiently large, $\mathcal{N}_1$ is positive, $\mathcal{N}_2$ negative, while when $\Ca$ is small $\mathcal{N}_1$ is actually negative and becomes positive only for $\Ca \approx 0.1$. Also, $\mathcal{N}_2$ assumes an opposite sign when the capillary number is small, in the limit of rigid particles. The change of sign of the normal stress differences marks the change of behaviour of the suspension, from a rigid-like behaviour dominated by particle-particle interactions to the deformable one dominated by the particle deformation and consequent tendency to migrate towards the centerline.

Finally, the total stress budgets revealed that the relative particle-induced stress contribution grows with the volume fraction and Reynolds number, and reduces as a function of the particle deformability. For all cases, since we are dealing with laminar suspension flows, the viscous stress is the dominant contribution and, as expected, is the only not vanishing component at the wall. It becomes smaller as the volume fraction is increased, but this is compensated by a larger particle stress contribution. Concerning the Reynolds stresses, it can be noted that they are almost negligible in every case under investigation.

The present analysis demonstrated the importance of studying suspension of deformable objects at finite inertia and for non-uniform shear rates. Indeed, we show that results can be different from what observed in Couette flows due to non-uniformity of the shear which, coupled with particle defomation, brings to a suspension behaviour which is dependent not only on the shear rate, but also on the local particle concentration. This should be carefully considered when proposing macroscopic models for suspensions of deformable objects.

\section*{Acknowledgments}
The authors acknowledge computer time provided by SNIC (Swedish National Infrastructure for Computing) and by the Scientific Computing section of Research Support Division at OIST.

\newcommand{\newblock}{}
\bibliographystyle{plain}

\end{document}